\documentclass[aps,twocolumn,prb,reprint,amsmath,amssymb,showpacs,superscriptaddress]{revtex4-2}

\usepackage{graphicx}
\usepackage{dcolumn}
\usepackage{bm}
\usepackage{amsmath,amsfonts,amsthm}
\usepackage{color}
\usepackage{mathrsfs}
\usepackage{float}
\usepackage{ulem}
\usepackage{txfonts}
\usepackage{wasysym}
\usepackage{cancel}

\newcommand{\MR}[1]{\textcolor{blue}{#1}}

\makeatletter

\newcommand{\Rmnum}[1]{\expandafter\@slowromancap\romannumeral #1@}
\makeatother
\begin{document}
\title{Josephson junctions of 2D time-reversal invariant superconductors: signatures of the topological phase}

\author{Gabriel F. Rodr\'{\i}guez Ruiz}
\affiliation{Escuela de Ciencia y Tecnolog\'{\i}a  and ICIFI, Universidad Nacional de San Mart\'{\i}n, Av 25 de Mayo y Francia, 1650 Buenos Aires, Argentina}
\author{Michael A. Rampp}
\affiliation{Institute for Theory of Condensed Matter, Karlsruhe Institute of Technology (KIT), 76131 Karlsruhe, Germany }
\affiliation{Max Planck Institute for the Physics of Complex Systems, 01187 Dresden, Germany}
\author{A. A. Aligia}
\affiliation{Instituto de Nanociencia y Nanotecnolog\'{\i}a CNEA-CONICET,
Centro At\'{o}mico Bariloche and Instituto Balseiro, 8400 Bariloche, Argentina}
\author{Joerg Schmalian}
\affiliation{Institute for Theory of Condensed Matter, Karlsruhe Institute of Technology (KIT), 76131 Karlsruhe, Germany }
\author{Liliana Arrachea}
\affiliation{Escuela de Ciencia y Tecnolog\'{\i}a  and ICIFI, Universidad Nacional de San Mart\'{\i}n-UNSAM, Av 25 de Mayo y Francia, 1650 Buenos Aires, Argentina}

\begin{abstract}
We determine the current-phase relation (CPR) of two-terminal configurations of Josephson junctions containing two-dimensional (2D) 
time-reversal invariant topological superconductors (TRITOPS), including TRITOPS-TRITOPS, as well as junctions between topological and non-topological superconductors (TRITOPS-S). 
We focus on wide junctions for which several channels intervene in the tunneling coupling. We  derive effective Hamiltonians to describe 
the topological edge modes for different TRITOPS models, including Hamiltonians with $p$-wave pairing and Hamiltonians combining $s$-wave pairing with spin-orbit coupling. We also derive effective low-energy Hamiltonians to describe the Josephson 
junction. These can be solved analytically and explain the contribution of the edge states to the Josephson current as a function of the phase bias. 
We find that edge-modes yield peculiar features to the CPR for both junction types. The primary effects occur for the response of the Majorana zero-modes at half-flux quantum phase $\phi\approx \pi$ in TRITOPS-TRITOPS junctions and for integer flux quantum phase $\phi \approx 0$ for TRITOPS-S junctions, respectively. The former effect is particularly strong for two-component nematic superconductors.
The second effect leads to a spontaneously broken time-reversal symmetry in the TRITOPS-S junction and to a breakdown of the bulk-boundary correspondence. 
We analyze in this case the role of the phase fluctuations. For weakly-coupled junctions, we show that time-reversal symmetry is restored for large enough stiffness in these fluctuations.
\end{abstract}

\date{\today}
\maketitle

\section{Introduction.} 
Topological superconductivity is among the most active research topics for some time now \cite{bernevig2013topological}. The topological superconductors are characterized by nontrivial topological quantum numbers in the bulk, which are usually  accompanied by subgap excitations 
localized at the edges that behave as gapless Majorana fermions. Those have  attracted  great  interest because of their potential application in quantum information processing \cite{kitaev2001unpaired,freedman2002modular,kitaev2003fault}. 

The simplest model for topological superconductivity is Kitaev's model, which  was formulated for spinless (or fully spin-polarized) fermions with $p$-wave pairing \cite{kitaev2001unpaired}. In 1D the subgap states are  Majorana bound states at zero energy that are localized at the ends of the superconducting wire. The latter  are represented by 
operators satisfying $\gamma^{\dagger} = \gamma$ and $\gamma^2=1$.  In 2D Majorana edge modes are massless and propagate along the edge in 1D channels satisfying 
$\eta^{\dagger}_k=\eta_{-k}$ and $\left\{\eta_{k} , \eta_{k^{\prime}} \right\}=\delta_{k,k^{\prime}}$. 
Such models guided the search for the topological phase  in more realistic systems, where singlet superconductivity is the dominant type. A promising platform for the realization of topological superconductivity 
is based upon the combination of $s$-wave singlet superconductivity with spin-orbit coupling (SOC) and magnetic fields, which effectively generates p-wave superconductivity \cite{wires1,wires2}. Several experiments in semiconducting wires with spin-orbit coupling in proximity with superconductors show features consistent with these ideas \cite{mourik2012signatures,rokhinson2012fractional,das2012zero,albrecht2016exponential,deng2012majorana}. Another avenue to engineer a 1D topological superconductor is based on magnetic adatoms inducing subgap states in superconducting substrates \cite{yazdani,wiesendanger,ruby}. Furthermore, the iron-based material FeSeTe, with intrinsic $s$-wave superconductivity and surface magnetism  \cite{iron} as well as topological insulators in proximity with ordinary superconductors and magnetic islands~\cite{fu2008superconducting,fu2009josephson} are also considered as a platform to realize Majorana states. 
Several results in this direction are reviewed in Refs. \onlinecite{zhang,alicea2012new,aguado2017majorana,flensberg2021engineered}.
All the systems mentioned above rely on mechanisms breaking time-reversal symmetry. On the basis of symmetry analysis, 
it was recognized early on that  other families of topological superconductors may exist \cite{ryu2010topological}.
 Those preserving time-reversal symmetry
are referred to as members of  the {\it DIII-class} or TRITOPS  (time-reversal symmetric topological superconductors). 
The key ingredient to realize this topological phase is the existence of two channels in which the pairing function have opposite signs\cite{qi2009time}. Formally,
a simple way to generate this effect is with two copies of Kitaev's model related by time-reversal symmetry \cite{dumitrescu2013topological,haim2014time,tanaka-tritops} or by considering  time-reversal-symmetric p-wave pairing \cite{kwon}.
Several theoretical proposals have been formulated in a number of systems. These include 2D and 3D models \cite{qi2009time,fu2010odd,deng2012majorana,scheurer2015topological}, as well as architectures of real systems like
nanowires with Rashba spin-orbit coupling with proximitized d-wave  \cite{wong2012majorana} or extended s-wave \cite{zhang2013time}, configurations of two wires with 
spin-obit coupling, s-wave superconductivity and magnetic fields in arrangements globally preserving time-reversal symmetry 
 \cite{keselman2013inducing,haim2014time,haim2016interaction,reeg2017diii}, 
  2D topological insulators in proximity with superconductors \cite{santos2010superconductivity,klinovaja2014kramers,mellars2016signatures,parhizgar2017highly,casas2019proximity}
and  thin films of iron-based superconductors \cite{zhang2021intrinsic}. As a consequence of the time-reversal symmetry, the edge modes of these topological systems appear in Kramers's pairs of Majorana modes. Their signatures can be identified in the noise spectrum and in the behavior of the Josephson 
current \cite{keselman2013inducing,chung2013time,nakosai2013majorana,schrade2015proximity,li2016detection,alicea,camjayi2017fractional,schrade2018parity,aligia2018entangled,haim2019time,gong2016influence,mashkoori2019impact,lauke2018friedel}.

The hybridization between the topological edge states of topological superconductors in a Josephson junction 
 leads to the formation of Andreev bound states. In 1D TRITOPS, the edge modes have zero energy and are localized at the end of the system.
 The corresponding Andreev bound states are characterized by symmetry-protected level crossings, which give rise to  jumps in the current-phase relation (CPR) $J\left(\phi \right)$, being $\phi$ the phase bias at the junction. Such features depend on the 
 structure of these modes, in particular on the spin projection of their particle and hole components
\cite{arrachea2019catalog,haim2019time,haim2019spontaneous,alicea}. In 2D, the edge modes extend along the boundaries of the system.
One of the goals of the present work is to analyze the structure of these modes, 
in particular, their dispersion relation and their spin structure.

In 2D, the characteristics of the Majorana edge modes are not universal but depend on the nature of the bulk. 
We show that  they depend, in particular, on the details of the pairing mechanism and are also affected by the presence of the spin-orbit 
coupling. To properly analyze and compare these effects, we do not restrict ourselves to a single type of 
TRITOPS but consider models with and without spin-orbit coupling.
We focus on two families of BCS models in 2D, which are representative of  the  different proposals reported in the literature: (a)  
{\it $p$-wave pairing}. Here our aim is to analyze the effect of spatial symmetry, which enables superconducting phases represented by one and two-dimensional order parameters. This is motivated by the observation of a nematic phase in the doped topological insulator Cu$_x$Bi$_2$Se$_3$\cite{matano2016spin,yonezawa2017thermodynamic}, which  has been suggested to be 
 a TRITOPS with odd-parity superconducting pairing \cite{fu2010odd}. 
(b) {\it $s_{\pm}$-wave pairing in combination with SOC}, which are the ingredients of the TRITOPS platforms based on unconventional superconductors \cite{zhang2013time,zhang2021intrinsic}.
We derive effective Hamiltonians for the edge modes in each case which we compare with numerical results. These consist of 1D Dirac Hamiltonians describing the dynamics of the Kramers pairs of Majorana modes. The  velocity of propagation of these modes as well as 
the structure of the spinors describing them are determined by the pairing mechanism of the bulk Hamiltonian and by the presence of the SOC.

The other  goal of the present work is to analyze the impact that the structure of the edge modes have on the behavior of the CPR. 
We consider two types of Josephson junctions: (i) TRITOPS-TRITOPS and (ii) TRITOPS-S (S denotes a conventional superconductor). 
We derive effective low-energy Hamiltonians for these configurations, which can be solved analytically. The coupling of the edge modes in the junction generate $\phi$-dependent mass terms in the Dirac Hamiltonians, which reveals the different nature of the junction. In the TRITOPS-TRITOPS case, the mass term is $\propto \cos(\phi/2)$ implying the opening of a gap in the spectrum of the topological Andreev modes close to $\phi=0$. This mass term depends on the type of pairing and may have a complex structure which depends on the SOC. 
Instead, in the TRITOPS-S case, the mass generation is much more subtle. In this case the junction separates phases of different topology and hence the bulk-boundary correspondence demands the edge to host gapless modes. This is reflected in the $\propto \sin(\phi)$ dependence of the mass term.
This is a consequence of the fact that  for $\phi=0$ the Kramers' pair of edge modes  remains robust under the coupling to the non-topological S system. However, as soon as the time-reversal symmetry is broken by a small $\phi$, a gap develops in the corresponding Andreev spectrum. 
We show that this mechanism is very general and it takes place irrespective of the details of the pairing mechanism and the SOC. The outcome is a jump of the CPR at $\phi=0$, implying an instability of the bulk-boundary correspondence as soon as the time reversal protecting symmetry is broken.
The work is organized as follows. We introduce the models to be investigated in Section II.  Section III is devoted to analyze the topological properties of the different models and to derive the effective Hamiltonians for the edge modes. We analyze the Josephson current in Section IV. Here we solve the problem numerically by diagonalizing exactly the lattice Hamiltonians and we also derive effective low-energy models based on the Josephson-tunneling coupling of the edge modes, which can be solved analytically. In all the cases we focus on junctions with many transverse channels that we analyze in the momentum space.  Section V is devoted to analyze in detail the instability of the 
TRITOPS-S junction. Section VI contains a summary and conclusions, some technical details are presented in Appendices A to D.

\section{Models for the TRITOPS phase}
We consider two different types of 2D models with BCS pairing defined in the square lattice and hosting a TRITOPS phase. (a) Models with $p$-wave pairing preserving time-reversal symmetry. The most studied case in the literature consists of two copies 
of the Kitaev model \cite{qi2009time,dumitrescu2013topological,haim2014time}, where each copy has triplet pairing of fully polarized fermions. However, this is not the only possibility, since it is also possible to have triplet $p$-wave pairing between electrons with opposite spin orientation as it is well known in the context of He$^3$\cite{vollhardt2013superfluid,read2000paired}. Taking also into account  the symmetry properties of the underlying lattice, we analyze the structure of the edge modes in the  different  irreducible representations of the $p$-wave pairing order parameter. This analysis is important in view of the nematic phase observed in the superconducting phase of the doped topological insulator Cu$_x$Bi$_2$Se$_3$\cite{matano2016spin,yonezawa2017thermodynamic}. Although this phase takes place in 3D, two-dimensional architectures based on this compound could inherit similar properties. 
We anticipate that, while the one-dimensional irreducible representations host dispersing edge modes, the edge modes 
of the
two-dimensional one are dispersionless. 
(b) We also study a model where the pairing is of extended $s$-wave type in combination with SOC. Here, we will see that the combination of these two ingredients effectively generates a $p$-wave type pairing in the one-dimensional irreducible representations of the 2D lattice but with a spin structure of the edge modes affected by the SOC. In the forthcoming sections, we will  analytically derive effective Hamiltonians for the edge modes and we will see how all these features lead to different signatures in the behavior of the CPR.

\subsection{$p$-wave  pairing}
We consider the following Hamiltonian in the lattice $H = \frac{1}{2}\sum_{\bf k} {\bf c}^{\dagger}_{\bf k} H^{p}_{\bf k}{\bf c}_{\bf k}$, with
${\bf c}_{\bf k}= \left(c_{\bf k, \uparrow},c_{\bf k, \downarrow}, c^{\dagger}_{\bf -k, \downarrow}, - c^{\dagger}_{\bf -k, \uparrow}\right)^T$ and ${\bf k}=\left( k_x,k_y \right)$, while the Bogoliubov de Gennes Hamiltonian matrix reads
\begin{equation}\label{h-p-pm}
    H^{p}_{\bf k}  =  \xi_{\bf k} \tau^z \sigma^0 +  \tau^x \boldsymbol{\sigma} \cdot \boldsymbol{\Delta}^{\alpha, \beta}_{\boldsymbol{k}}.
\end{equation}
The Pauli matrices $\tau^{x,y,z}$ and $\boldsymbol{\sigma}=(\sigma^{x},\sigma^y,\sigma^z)$ act, respectively,  on the particle-hole and spin degrees of freedom,
while $\tau^0, \sigma^0$ are $2\times 2$ identity matrices. The dispersion relation is defined in terms of a hopping element $t$ as  $\varepsilon_{\bf k}=- 2 t \left(\cos k_x + \cos k_y \right)$, 
hence $\xi_{\bf k}=\varepsilon_{\bf k} -\mu $, being $\mu$ the chemical potential. Our results do not rely on the assumption of only nearest neighbor hopping in $\varepsilon_{\bf k}$ and easily carry over to dispersions with further range hoppings.
The $p$-wave pairing vector function, restricting to a ${\bf k}$ dependence with only linear terms in $\sin k_x$ and $\sin k_y$ and preserving time-reversal symmetry, reads
\begin{equation}
\boldsymbol{\Delta}^{\alpha, \beta}_{\boldsymbol{k}} =  \Delta_x \sin k_x \;{\bf n}^{\alpha}+ \Delta_y \sin k_y \;{\bf n}^{\beta},
\end{equation}
with $(\Delta_y, \Delta_y)$ real.
With the above restrictions, it is possible to build a pairing vector function 
for each of the irreducible representations of the point group $D_{4h}$,
\begin{eqnarray}\label{deltap}
   \boldsymbol{\Delta}^{{\rm A}_{1u}}_{\boldsymbol{k}} &= & \Delta \left( \sin k_x \;{\bf n}^x+  \sin k_y \;{\bf n}^y \right) \nonumber \\
    \boldsymbol{\Delta}^{{\rm A}_{2u}}_{\boldsymbol{k}} &= & 
    \Delta \left(\sin k_y \;{\bf n}^x- \sin k_x \;{\bf n}^y \right), \nonumber \\
    \boldsymbol{\Delta}^{{\rm B}_{1u}}_{\boldsymbol{k}} &= & \Delta \left(\sin k_x \;{\bf n}^x -  \sin k_y \;{\bf n}^y\right),\nonumber \\
     \boldsymbol{\Delta}^{{\rm B}_{2u}}_{\boldsymbol{k}} &= & \Delta \left( \sin k_x \; {\bf n}^y +  \sin k_y \; {\bf n}^x \right)\;,\nonumber \\
      \boldsymbol{\Delta}^{{\rm E}_{u}}_{\boldsymbol{k}} &= & \Delta \left( \sin k_x \pm  \sin k_y \right)\;{\bf n}^z,\nonumber \\
\end{eqnarray}
 ${\bf n}^{x,y,x}$ being unit vectors along the $x,y,z$-directions. The A$_{ju}$, B$_{ju}$, $j=1,2$ are one-dimensional irreducible representations, while the E$_u$ is
 two-dimensional.
For an intrinsic superconductor, the allowed values for the two components $\left(\Delta_{x},\Delta_{y}\right)$ are determined
by the non-linear, quartic terms of the Ginzburg-Landau expansion.
The three options that result are, on the one hand
two solution proportional to either $\left(1,\pm1\right)$ or $\left(1,0\right)$
and $\left(0,1\right)$. Those are nematic superconductors where the
superconducting state breaks a rotation symmetry. On the other hand, there
is the option proportional to $\left(1,\pm i\right)$ which breaks
time-reversal symmetry and hence is not of the type discussed in this paper. Alternatively, superconductivity could be
the consequence of a proximity effect to a substrate, in which case 
all the real combinations of $\left(\Delta_{x},\Delta_y\right)$ are consistent with the time-reversal symmetry.
In addition, the edges 
of the samples are not necessarily aligned with the crystalline axes. 
We consider the particular real nematic phase with $\boldsymbol{\Delta}^{{\rm E}_{u}}_{\boldsymbol{k}}$ defined in Eq. (\ref{deltap}) 
but we have checked that our conclusions hold for any other choice of $(\Delta_x, \Delta_y)$.

\subsection{$s_{\pm}$-wave  pairing and SOC}
The second type of model we will analyze is based on BCS pairing with $s$-wave symmetry in combination with spin-orbit coupling. We focus, in particular, on the model proposed by Zhang-Kane-Mele (ZKM) in Ref. \onlinecite{zhang2013time}, which is a BCS Hamiltonian with local $\Delta_0$ plus extended $\Delta_1$ s-wave pairing and Rashba spin-orbit coupling (SOC) $\lambda$. 
The Hamiltonian reads
\begin{eqnarray}\label{h-zkm}
 H^{\rm ZKM}_{\bf k}  &=&  \;   \xi_{\bf k} \tau^z \sigma^0 + 2 \lambda \tau^z \left(\sin k_x  \sigma^y - \sin k_y \sigma^x \right)  \nonumber \\
 & &  + \; \tau^x \sigma^{0	}\Delta_{\bf k}.
\end{eqnarray}
The pairing potential has a local $\Delta_0$ plus an extended $\Delta_1$ s-wave components,
with $\Delta_{\bf k}=\Delta_0 + 2 \Delta_1\left(  \cos k_x + \cos k_y \right)$.
This model hosts a topological phase  for $|\mu-\varepsilon_0| < \varepsilon_{1}$, with $\varepsilon_0=t \Delta_0/\Delta_1$ and
 $\varepsilon_1= 2 \lambda \sqrt{| \Delta_0/\Delta_1| - \Delta_0^2/(4 \Delta_1^2)}$. 
 
 \section{Effective Hamiltonians for the edge modes}
 The TRITOPS phase is characterized by the existence of  Kramers' pairs of  Majorana edge  modes. 
 The aim of this section is to derive effective Hamiltonians to analytically describe the dynamics of these modes. This will be the starting point to analytically describe the  Andreev spectra generated when these states are coupled in the Josephson junction. We focus on the two families of models previously introduced. 
 For simplicity,  we start the discussion with an analytic investigation of edge modes in the continuum limit. In the case of the ZKM model we must rely on 
 an analytical solution of the lattice model in order to capture all the details introduced by the SOC. In all the cases we compare with the solution of the lattice Hamiltonian with a numerical approach.

\begin{figure}[h]
\begin{center}
\includegraphics[width=\columnwidth, trim={0 0cm 0 0cm}, clip]{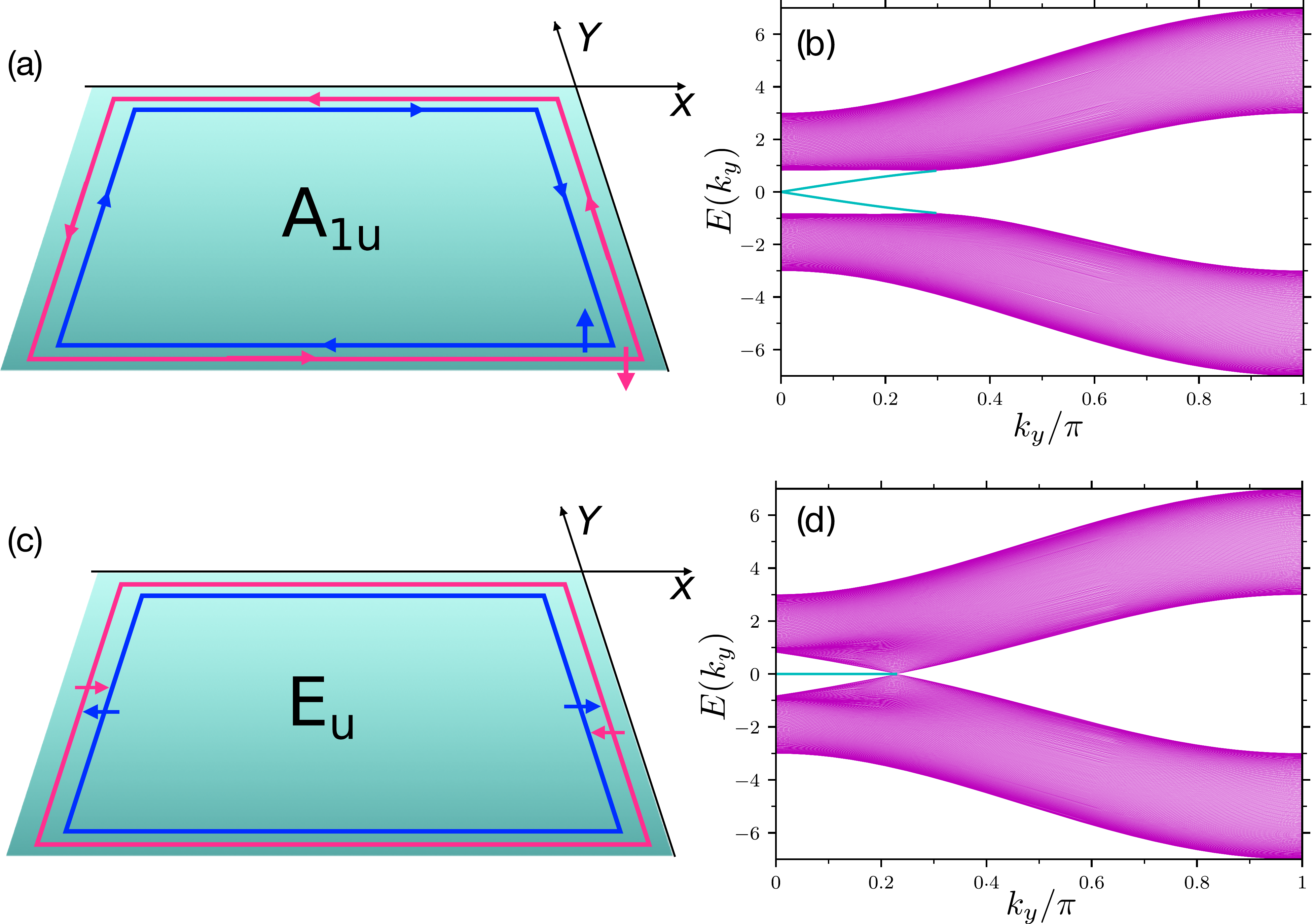}
\end{center}
\caption{Edge states and spectra for the topological phase of Hamiltonians with $p$-wave pairing. (a) and (b) correspond to the Hamiltonian of Eq. (\ref{h-p-pm}), with $\Delta_{\bf k}$ belonging to the one-dimensional irreducible representations of
Eq. (\ref{deltap}), while (c) and (d) to the two dimensional representation E$_u$. Only A$_{1u}$ (equal to A$_{2u}$)
is shown. B$_{1u}$ (equal to B$_{2u}$) has the same spin structure with opposite chiralities. The spectra are calculated for a system with open boundaries along the $x$ direction and periodic boundary conditions along $y$
(only $k_y\geq 0$ is shown). The edge states are indicated in light blue. These are two-fold and four-fold degenerate in (b) and (d), respectively.}
\label{fig1}
\end{figure}
 
 \subsection{$p$-wave  model with A$_{ju}$ and B$_{ju}$ symmetry}\label{pab}
 A simple derivation of the effective Hamiltonian to describe  the  edge modes is possible by considering the continuum version of the Hamiltonian of 
 Eq. (\ref{h-p-pm}). We start by analyzing the cases with $\Delta_{\bf k}^{{\rm A}_{1u}}$ and  $\Delta_{\bf k}^{{\rm B}_{1u}}$, which corresponds to
  \begin{equation}
     {\cal H}^{(1)}=\frac{1}{2}\int d^2x \; \Psi^{\dagger}({\bf x})\; H^{(1)} \; \Psi({\bf x}),
 \end{equation}
 with the Bogoliubov-deGennes Hamiltonian
 \begin{equation}\label{h-p-pm-cont}
H^{(1)}= \tau^z \sigma^0 \left[\varepsilon_{\bf p}-\mu(x)\right]+\Delta \tau^x \left(p_x \sigma^x \pm p_y \sigma^y \right),
\end{equation}
where $\pm$ corresponds to A$_{1u}$ and B$_{1u}$, respectively.
The Nambu field operators are defined as $\Psi({\bf x})=\left(\psi({\bf x}), i \sigma^y \psi^{\dagger}({\bf x}) \right)^T$,  being $\psi({\bf x})=(\psi_{\uparrow}({\bf x}), \psi_{\downarrow}({\bf x}))^T$ a spinor in  spin space, while
${\bf x}=(x,y)$ and $p_x,p_y$ denote the momentum in the $x$ and $y$ direction, respectively, with the dispersion relation $\varepsilon_{\bf p}=p^2/(2m)$.

The  Hamiltonians for the edges along $y$ read (see Appendix \ref{papp} for details)
\begin{equation}
    H^{\nu}=\sum_{p_y \geq 0,\sigma}{\rm v}_{\nu,\sigma}p_y \eta_{\nu, p_y,\sigma}^{\dagger} \eta_{\nu,p_y,\sigma},
\end{equation}
where $\nu=l,r$ labels the left of right edges of a long ribbon along the $y$ direction and  ${\rm v}_{\nu,\sigma}= s_{\nu} s_{\sigma} \Delta$ is the velocity of propagation of the modes, with $s_{\uparrow}=-s_{\downarrow}=1$. 
The corresponding Bogoliubov  operators are 
\begin{equation} \label{gammanu}
\eta_{\nu,p_y,\sigma}=\frac{e^{is_{\nu} s_{\sigma} \pi/4}}{\sqrt{2}} \left(c_{\nu,p_y,\sigma}-i s_{\nu} s_{\sigma} c^{\dagger}_{\nu,-p_y,\sigma}\right),
\end{equation}
where $c_{\nu,p_y,\sigma}$ is the annihilation operator of a fermion with momentum $p_y$ and spin $\sigma$ at the edge $\nu$.
Notice that the Bogoliubov operators describing the edge modes, given in Eq. (\ref{gammanu}), satisfy 
the condition
\begin{equation}
    \eta_{\nu,p_y,\sigma}^{\dagger}= \eta_{\nu,-p_y,\sigma}.
\end{equation}
The solution for the edges along the $x$-direction is similar and the picture is consistent with two helical Majorana modes with associated opposite spin orientations circulating along the edges with opposite chiralities (see sketch of Fig. 1.a). The corresponding spectrum is presented in Fig. 1.b. 
The analysis of the representations A$_{2u}$ and B$_{2u}$ is completely analogous and the solution is the same with an identical result.

\subsection{p-wave model with E$_u$ symmetry}
We can proceed in a similar way as in Sec. \ref{pab}. The Bogoliubov-De-Gennes Hamiltonian for the continuum version in the present case reads
\begin{equation}\label{h-p-0-cont}
H^{{\rm E}_u}= \tau^z \sigma^0 \left[\varepsilon_{\bf p}-\mu(x)\right]+\Delta \tau^x \sigma^z \left(p_x \pm p_y \right),
\end{equation}
where, as before, we consider $\Delta >0$ and the topological phase corresponds to $\mu>0$. 
The calculation of the zero modes for $p_y=0$ leads to a solution with identical structure as Eq. (\ref{zero-p-z}), but
with $\Lambda^{\nu}_{0, s}$ being a spinor that satisfies
$\tau^y \sigma^z \Lambda^{\nu}_{0 s}=s_{\nu} \Lambda^{\nu}_{0 s}$, with $s_r=-s_l=1$. Hence, $\Lambda^{\nu}_{0 +}=\frac{1}{2}\left(1,1,s_{\nu}i,-s_{\nu}i \right)^T$
and $\Lambda^{\nu}_{0 -}=\frac{1}{2}\left(1,-1,s_{\nu}i,s_{\nu}i \right)^T$. Remarkably, the solution for $p_y\neq 0$ corresponds to evanescent modes, which is consistent with a flat band of  zero modes localized at the edges. Therefore, the edge modes are  non-dispersive. The sketch of these states along with the spectrum is shown in Fig. 1.c and 1.d, respectively.

\subsection{ZKM model}
 
 \subsubsection{Simplified continuum version}
\begin{figure}[h]
\begin{center}
\includegraphics[width=\columnwidth]{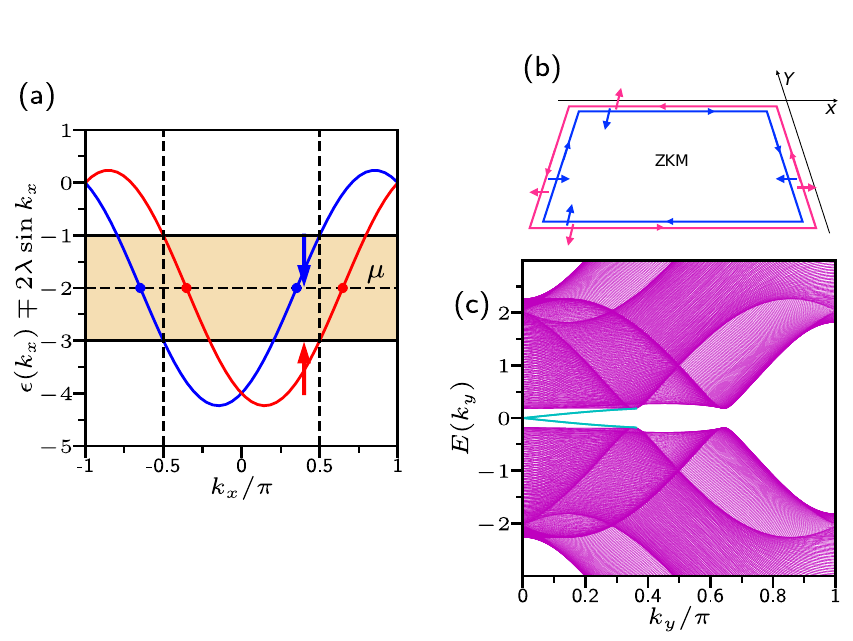} 
\end{center}
\caption{(a) Bands of the Hamiltonian $H^{\rm ZKM}$ without pairing for $k_y=0$. The yellow region indicates the range of values of $\mu$ within which the topological phase develops. The Fermi points indicated in dots are $-k_{F1},\;-k_{F2},\;
k_{F2},\;k_{F1}$ (from left to right). (b) Sketch of the edge states for the continuum Hamiltonian. (c) Spectrum of $H^{\rm ZKM}$ with periodic boundary conditions along $y$ and open boundary conditions along $x$ in a system with $N_x=200$. Parameters are
$\Delta_0=-2\Delta_1=-0.4t$, $\lambda=0.5t$ and $\mu=-2t$. The (doubly degenerate)} edge states are indicated in light blue.
\label{fig:2}
\end{figure}

 To proceed as in the case of the $p$-wave BCS model, we define a low-energy continuum Hamiltonian for the lattice model defined in Eq. (\ref{h-zkm}). 
The pairing potential of this model has a nodal surface for which $\Delta_k=0$, which encloses the time-reversal-invariant point ${\bf k}_0=(0,0)$ for $\Delta_0/\Delta_1 <0$,  
or ${\bf k}_0=( \pi,  \pi)$ for $\Delta_0/\Delta_1 >0$ \cite{zhang2013time} and the topological phase develops when the Fermi energy approaches this surface. Due to the SOC, the system without pairing has two bands with different Fermi surfaces. The dispersion relation for $k_y=0$ is shown in
Fig. \ref{fig:2} a. 
The 
continuum model is obtained by linearizing this Hamiltonian with respect to the Fermi points of these two bands at the Fermi energy of the nodal surface of $\Delta_{\bf k}$.  The procedure is explained in Appendix \ref{zkm-cont}. The effective low energy Hamiltonian has  $p$-wave pairing in the representations A$_{ju}$ or B$_{ju}$ along with SOC as an additional ingredient.

The Hamiltonian for the $y$-edges reads
\begin{equation}\label{edge-zkm}
    H^{\nu}=\sum_{p_y \geq 0,s=\pm}{\rm v}_{s} p_y \eta_{p_y,s}^{\dagger} \eta_{\nu,p_y,s},
\end{equation}
with ${\rm v}_{s}=s 2 \lambda$.
Similar to the case of Eq. (\ref{gammanu}), the Bogoliubov operators describing the edge modes, given in Eq. (\ref{gammanux}), satisfy 
the condition
\begin{equation}\label{majo}
    \eta_{\nu,p_y,s}^{\dagger}= \eta_{\nu,-p_y,s}.
\end{equation}
Notice, however, that the spin orientation is along the $x$-direction in the present case.
 The solution for the edges running along the $x$-direction is similar but with the spin orientation along $y$.  The picture is consistent with two helical Majorana modes circulating along the edges with opposite chiralities and the spin texture shown in the  sketch of Fig. \ref{fig:2}.b. This is consistent with the   spectrum 
 calculated by the exact diagonalization of Eq. (\ref{h-zkm}), which is presented in Fig. \ref{fig:2}.c. 
  

\subsubsection{General solution in the lattice model}
In the previous analysis we have linearized the Hamiltonian with respect to  ${\bf k}$-points with one component of ${\bf k}_0$ kept fixed and the other component on the nodal lines of $\Delta_{\bf k}$ and we have calculated the corresponding effective Hamiltonians for the edge modes. For sake of simplicity, we have neglected low-energy terms corresponding to linearizing the Hamiltonian with respect to other  ${\bf k}$-values of the 2D Fermi surface. In those cases, the dispersion relation with respect to $k_x$ keeping $k_y$ fixed is similar to the one shown in Fig. \ref{fig:2}.b, but with the orientation of the spin tilted with respect to $z$.  In order to account for such  more general context, we propose an ansatz for the description of the edge modes in terms of Bogoliubov operators
with the structure of Eq. (\ref{gammanux}) but with fermions having a tilted spin orientation. It reads
\begin{equation}\label{etaintro}
  \eta_{\nu,k,s}= \frac{e^{-is s_{\nu} \pi/4}}{\sqrt{2}} \left(\tilde{f}_{\nu,k,s} + i 
    s s_{\nu} \tilde{f}^{\dagger}_{\nu,-k,s}\right), ~~s=\pm
\end{equation} 
where $k$ denotes the transverse direction to the finite-length ribbon, along which the edge localizes.
The fermionic operators $\tilde{f}_{\nu,k,s}$ are
\begin{eqnarray}\label{fermions}
    \tilde{f}_{\nu,k,+} &=& e^{-i \delta_{\nu,k}} \left[ \cos(\frac{\theta_{\nu,k}}{2}) f_{\nu,k,\uparrow} + e^{-i \varphi_{\nu,k}} \sin(\frac{\theta_{\nu,k}}{2}) f_{\nu,k,\downarrow}\right]
\nonumber \\
 \tilde{f}_{\nu,k,-} &=& e^{i \delta_{\nu,k}} \left[ - e^{i \varphi_{\nu,k}} \sin(\frac{\theta_{\nu,k}}{2}) f_{\nu,k,\uparrow} +  \cos(\frac{\theta_{\nu,k}}{2}) f_{\nu,k,\downarrow}\right],
\end{eqnarray}
with $\theta_{\nu,k}=\theta_{\nu,-k}$, $\varphi_{\nu,k}=\varphi_{\nu,-k}$, $\delta_{\nu,k}=\delta_{\nu,-k}$, so that they
are time-reversal partners, ${\cal T} \tilde{f}_{\nu,k,+} {\cal T}^{-1}= \tilde{f}_{\nu,-k,-},~{\cal T} \tilde{f}_{\nu,k,-} {\cal T}^{-1}= -\tilde{f}_{\nu,-k,+}$. These operators describe localized fermions at the $\nu$ edge with spin-1/2 orientations along $\vec{n}_{\nu,k}=(\cos\theta_{\nu,k} \cos \varphi_{\nu,k}, \cos\theta_{\nu,k} \sin \varphi_{\nu,k}, \sin \theta_{\nu,k})$ in the coordinate system indicated in Fig. \ref{fig:2}. 
For this reason, the fermions of Eq. (\ref{fermions}) are basically the fermions $c_{\nu,p_y,s}$ of Eq. (\ref{gammanux}) upon applying a SU(2) operation that tilts the spin from an orientation along the $x$-direction
to  $\vec{n}_{\nu,k}$. Following the reasoning of Ref. \onlinecite{aligia2020tomography}, we notice that a SU(2) rotation in the fermions defining Majorana modes comes along with a change in the phases. For this reason,
we introduced the phase  $\delta_{\nu,k}$ in Eq. (\ref{fermions}), which, together with $\varphi_{\nu,k}, \theta_{\nu,k}$, define the generalized Bloch coordinates for each $k$-value along the edge state.

This heuristic argument can be verified by following a similar procedure as in Refs. \onlinecite{alaseprl,alaseprb}, modified to get  analytical results as explained in  Appendix \ref{ex-sol}.
Concretely, we consider the following lattice Hamiltonian, 
\begin{eqnarray}\label{h-zkm-1}
  & &H^{\rm ZKM}_{k}  = \;   \sum_{j=1}^{L_x}
 {\bf c}^{\dagger}_{j,k}\left[
\tau^z \left(\xi_{k} - 2 \lambda \sin k  \sigma^x\right) + \Delta_k \tau^x 
\right]{\bf c}_{j,k} 
\nonumber \\
& & + \sum_{j=1}^{L_x-1} \left( {\bf c}^{\dagger}_{j,k} \left[  \tau^z\left(-t - i \lambda \sigma^z\right)  + \Delta_1 \tau^x \right]  
{\bf c}_{j+1,k}  + H.c. \right)
\end{eqnarray}
with ${\bf c}_{j,k}=\left(c_{j,k,\uparrow}, c_{j,k,\downarrow}, c^{\dagger}_{j,-k,\downarrow}, -c^{\dagger}_{j,-k,\uparrow}\right)^T$,
$\xi_k= -2 t \cos k -\mu$, $\Delta_k=\Delta_0 +2 \Delta_1 \cos k$. This corresponds to the Hamiltonian of Eq. (\ref{h-zkm}) defined in a slab of $L_x$ sites in the $x$-direction and periodic boundary conditions in the transverse $y$-direction (we are simplifying notation $k_y \equiv k$).   The solution in the neigborhood of $k=k_{0,y}$ is given by Eqs. (\ref{etaintro}) and (\ref{fermions}) with the angles
  $\theta_{\nu,k}=\pi/2$, $\varphi_{\nu,k}=-s_{\nu}\varphi_k$ and the phase 
$\delta_{\nu,k}=s_{\nu}\varphi_k/2$ with
\begin{equation}
 s_l=-s_r=\mbox{sign}(\lambda \Delta_1). 
 \end{equation}
 Hence, all the angles and phases of the generalized Bloch coordinates can be expressed in the present case in terms of a single $k$-dependent phase $\varphi_k$.
The fermionic operators $f_{\nu,k,\sigma}$ are related to the fermionic operators of the basis of the lattice model as follows,
\begin{eqnarray}\label{fnu}
 f_{l,k,\sigma} &=& {\cal N}_k \sum_{\ell=1}^2\alpha_{k,\ell,\sigma} \sum_{j=1}^{L_x} z_{k,\ell,\sigma}^{j-1}  c_{k j \sigma},\nonumber \\
  f_{r,k,\sigma}&=& {\cal N}_k \sum_{\ell=1}^2 \overline{\alpha}_{k,\ell,\sigma} \sum_{j=1}^{L_x} \overline{z}_{k,\ell,\sigma}^{L_x-j} c_{k j \sigma},
 \end{eqnarray}
 where
 ${\cal N}_k$ is a normalization factor, while $\alpha_{k,\ell,\uparrow}= \overline{\alpha}_{k,\ell,\downarrow}\equiv \alpha_{k,\ell}=\alpha_{-k,\ell}$ and $z_{k,\ell,\uparrow}=\overline{z}_{k,\ell,\downarrow}
 \equiv z_{k,\ell}=z_{-k,\ell}$ are complex coefficients which are determined by the open boundary conditions. 
 The Hamiltonian for the edge modes reads
 \begin{equation}\label{edge-latt-1}
H_{\nu}=\sum_{k > 0 , s=\pm} s \varepsilon_{\lambda,k} ~\eta_{\nu,k,s}^{\dagger} \eta_{\nu,k,s},
\end{equation}
with
\begin{equation} \label{ek}
 \varepsilon_{\lambda,k}=- 2  \rho_k \lambda \sin k. 
 \end{equation}
 The parameters $\rho_k$ and $\varphi_k$ are related to the parameters 
 $\alpha_\ell$ and $z_j$ through 
 \begin{equation}\label{rhok}
     \rho_k e^{i \varphi_k} = {\cal N}_k^2 \sum_{j=1}^{L_x}\left( \sum_{\ell=1}^2 \alpha_{k,\ell} z_{k,\ell}^{j-1} \right)^2.
 \end{equation}
 Importantly, $\rho_k\simeq \rho$, and $\varphi_k\simeq \varphi$ are approximately constant close to the Dirac point $k_{0,y}$, while $\rho_k$ tends to zero as $k$ significaly departs from this point.
The structure of the edge modes corresponds to the sketch of Fig. \ref{fig:2}, but with the spins tilted an angle $\varphi$ with respect to the plane of the superconductor.

\section{Josephson junction and CPR}

Our goal now is to analyze of the impact on the Josephson current of the different types of edge states corresponding to the different platforms for 
realizing the TRITOPS phase. To this end, we consider two superconductors contacted in a Josephson junction. The hybridization of the states of the two superconductors  leads to the development of Andreev states with energies below the superconducting gap. In the topological phase, these states are mainly originated by the hybridization between the edge states, which leads to peculiar features in the CPR. We analyze junctions between two TRITOPS as well as junctions between TRITOPS and an ordinary superconducting phase (S).

The Hamiltonian for the full system containing the two superconductors, ${\rm S}_1,\; {\rm S}_2$ and the tunneling junction is expressed as
$H= \sum_k H_k$ with 
\begin{equation}\label{hkfull}
H_k=\sum_{\alpha={\rm S_1,S_2}} H_{\alpha,k} + H_{{\rm J},k}.
\end{equation}
The Hamiltonian $H_{\alpha,k}$ corresponds to the TRITOPS Hamiltonian expressed in a slab of length $N_x$ and periodic boundary conditions in the transverse direction, adopting a representation as in Eq. (\ref{h-zkm-1}). 
The Hamiltonian for the tunneling junction is $H_{\rm J} =\sum_k H_{{\rm J},k}$, with
\begin{equation}
H_{\mathrm{J},k}=t_{\mathrm{J}}\sum_{\sigma}\left( e^{i\phi /2} c_{\mathrm{S1}
,k,1\sigma}^{\dagger } c_{\mathrm{S2},k,1,\sigma}+\text{H.c.}\right) ,  \label{junction}
\end{equation}
where $c_{\mathrm{S1},k,1,\sigma}^{\dagger }$ ($c_{\mathrm{S2},k,1, \sigma}^{\dagger }$) creates an electron with spin $\sigma$ in the
superconductor $\mathrm{S1}$  ($\mathrm{S2}$) at the boundary contacting the junction 
with wave vector $k$ in the transverse direction. The phase bias at the junction,
$\phi=2\pi \Phi /\Phi _{0}$, is defined by the total magnetic flux
$\Phi $, being
$\Phi _{0}=h/2e$  the flux quantum. Our aim is to analyze features originated in the intrinsic properties of the topological edge states. For this reason we focus on Josephson junctions without spin-orbit
effects. The latter usually introduce extra phases which affect the behavior of the Josephson current \cite{haim2019spontaneous}.

We calculate the Josephson current by diagonalizing exactly $H_k$ and evaluating the energy of the ground state of this many-body Hamiltonian
as \cite{aligia2020tomography, aligia2019}
\begin{equation}
 E_{0}(k,\phi)=-\frac{1}{2}\sum_{s=\pm} \varepsilon_{k,s}(\phi), \;\;\;\;\;\;\; J(k,\phi)= \frac{2e}{\hbar}\frac{\partial E_{0}(k,\phi)}{\partial \phi}.
 \end{equation}
The energies $\varepsilon_{k,s}(\phi)$ are the negative single-particle energies of $H_k$. The total Josephson current as a function of $\phi$
(CPR) is simply calculated as $J(\phi)=\sum_k J_k(\phi)$.
In all the cases, we compare the exact numerical results with analytical ones that are obtained by substituting the exact Hamiltonians for the superconductors by effective Hamiltonians representing only the edge modes of the TRITOPS and/or a simplified version of the ordinary superconductor.

\subsection{TRITOPS-TRITOPS junction}

\begin{figure}[h]
\begin{center}
\includegraphics[width=\columnwidth]{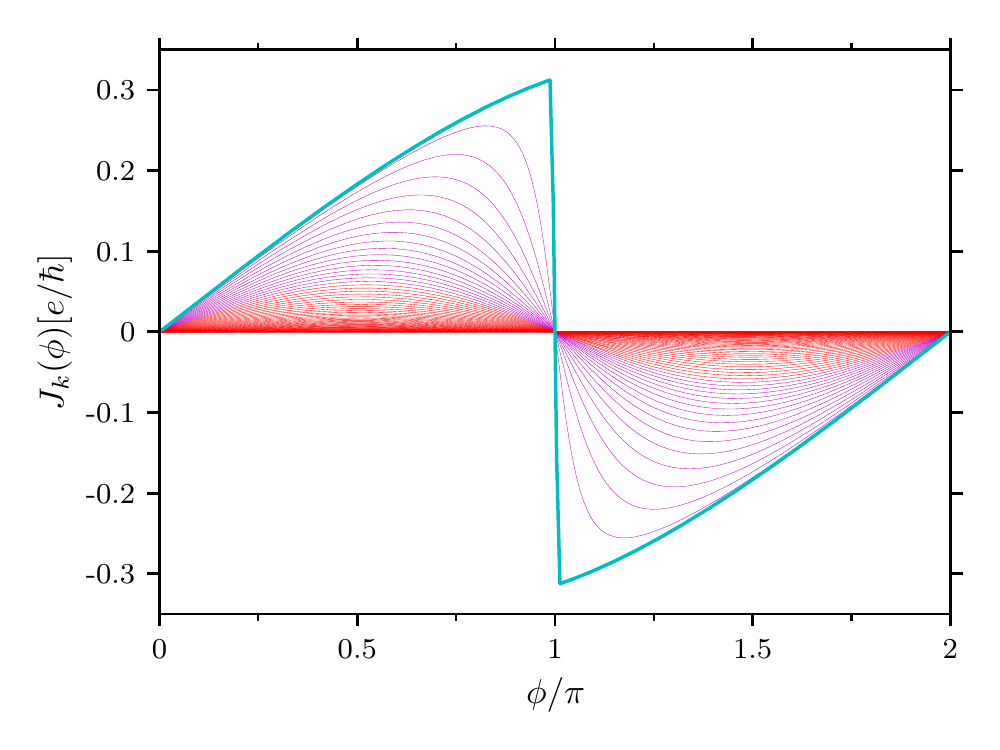}
\includegraphics[width=\columnwidth]{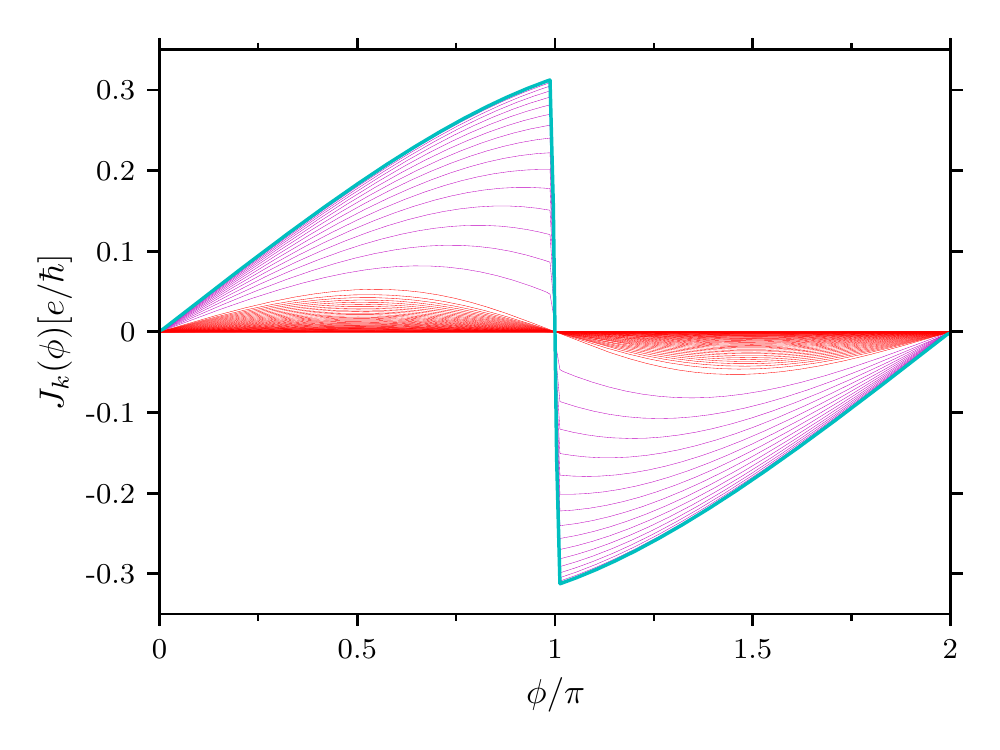}
\end{center}
\caption{$k$-resolved Josephson current (in units of  $e/\hbar$) as a function of the phase difference in the topological phase for a junction of 2D topological superconductors with $p$-wave pairing calculated with numerical exact diagonalization.  The upper and lower panel corresponds to the representations A$_{1u}$ (identical results are obtained for the A$_{2u}$ and B$_{ju},\; j=1,2$) and $E_u$, respectively. The plots in thick lines corresponds to $k=0$. Plots in violet correspond to the edge states, while the other 
$k$ values are shown in red. The parameters are $t_J=\lambda=0.5t$, $\Delta=t$ and $\mu=-3t$. A similar behavior is observed for other parameters within the topological phase ($-4t\leq\mu\leq 4t$)}
\label{figpk}
\end{figure}

\subsubsection{$p$-wave model}
Results for the Josephson current for different $k$-values in junctions between  TRITOPS with $p$-wave pairing are shown in Fig. \ref{figpk}. The two panels of the figure illustrate the behavior of this quantity for the different representations of the $p$-wave pairing introduced before. The different colors distinguish the contributions associated to the hybridization of the edge modes from those corresponding to the hybridization of the continuum states. 
We can see the impact of the different structure of edge modes in the two cases. 

\begin{figure}[h]
\begin{center}
\includegraphics[width=\columnwidth]{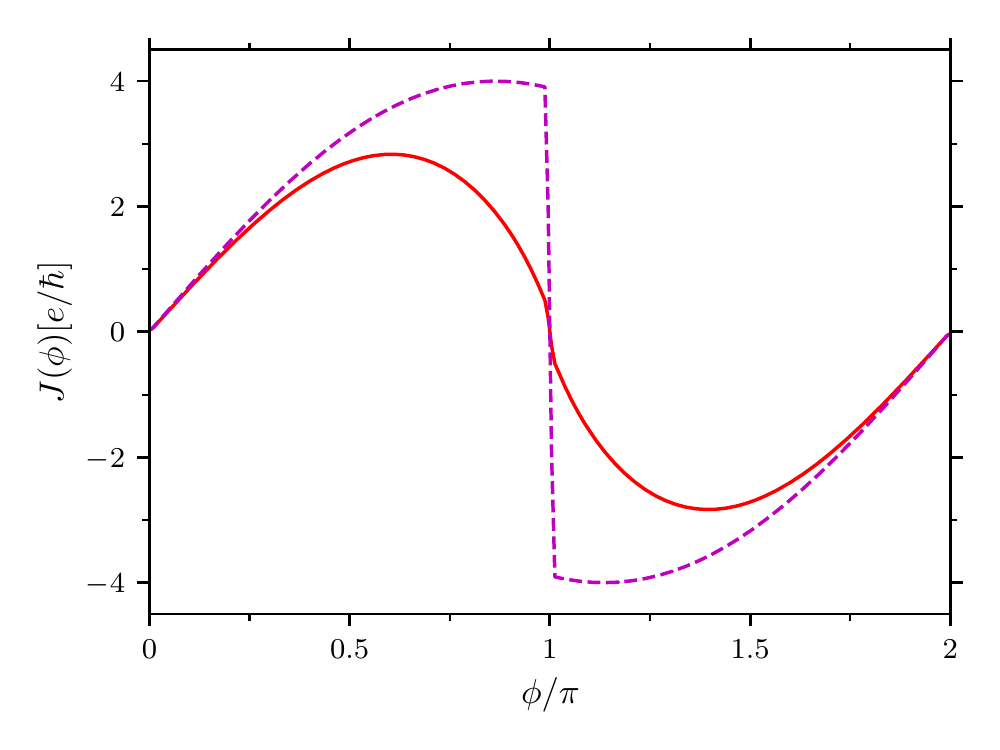}
\end{center}
\caption{Josephson current (in units of  $e/\hbar$) as a function of the phase difference in the topological phase for a junction of 2D topological superconductors with p-wave pairing
calculated by numerically diagonalyzing the coupled lattice Hamiltonians.  
Solid and dashed lines 
correspond to the representation $B_{1u}$ and $E_u$ for $t_J=\lambda=0.5t$, $\Delta=t$ and $\mu=-3t$.}
\label{figp}
\end{figure}

The B$_{1u}$ case is shown in the upper panel and we recall that the spectrum of the edge modes has a linear dispersion relation. The contribution of the zero-mode leads to a Josephson current which has a discontinuity at $\phi=\pi$ (see light-blue plot). This is the same behavior observed in topological superconducting wires and is a consequence of a level crossing of the Andreev states resulting from the hybridization of the Majorana zero-modes \cite{kwon,fu2009josephson,haim2014time,arrachea2019catalog,haim2019time}. Instead, the Josephson current is continuous as a function of $\phi$ for all the other edge modes with finite energy. Nevertheless, the observed behavior differs from the usual
$\propto \sin (\phi)$ function of non-topological junctions (see violet plots). The latter behavior is observed only for $k$-values associated to the continuum states (see red plots). We will see below that the Josephson coupling introduces a mass term in the effective Dirac Hamiltonian describing the free edge states which explains the peculiar CPR of the propagating Majorana edge states. 
Identical results are obtained for
the representations A$_{2u}$ and B$_{ju},\; j=1,2$.
For the E$_u$ case, where the edge modes form 
a flat band at zero energy, not only the $k=0$-mode but all the edge modes show a discontinuity at $\phi=\pi$ (see lower panel of Fig. \ref{figpk}).
The CPR is shown in Fig. \ref{figp} and is a superposition of all the $k$-components.

In both types of junctions, the behavior of the  Josephson current for $k$-values associated to the edge states can be explained in terms a low-energy effective Hamiltonian for the junction \cite{camjayi2017fractional,arrachea2019catalog,aligia2020tomography}, where we replace the fermionic operators for the TRITOPS' boundaries close to the junction 
$c_{\mathrm{S1},k,1,\sigma}\equiv c_{r,k,\sigma}$ and 
$c_{\mathrm{S2},k,1,\sigma}\equiv c_{l,k,\sigma}$ in Eq. (\ref{junction}) 
by their projections on the low-energy subgap excitations given by Eq. (\ref{gammanu}). Concretely, we perform the gauge transformation $ \eta^{\dagger}_{\nu,k,\sigma} \rightarrow e^{is_\nu s_{\sigma} \pi/4} \eta^{\dagger}_{\nu,k,\sigma}$ and we substitute
  \begin{equation}\label{ceffp}
 c^{\dagger}_{\nu,k,\sigma} \simeq  \frac{1}{\sqrt{2}} \eta^{\dagger}_{\nu,k,\sigma}, \nonumber \\
 \;\;\;\;\;\;
 c^{\dagger}_{\nu,-k,\sigma}  \simeq  \frac{is_\nu s_{\sigma} }{\sqrt{2}} \eta_{\nu,k,\sigma}.
  \end{equation}
  
  
  Introducing $\eta_{\nu,k}=(\eta_{\nu,k,\uparrow}, \eta_{\nu,k,\downarrow})^T$, for $k \geq 0$, we get 
  the following effective Hamiltonian for the junction, obtained after adding the contributions of $k$ and $-k$ in the original Hamiltonian
\begin{equation}\label{hkeff-tt-p}
  H^{p-p }_{ {\rm eff}, k} = t_{\rm J} \cos (\phi/2) \eta^{\dagger}_{l,k} \eta_{r,k} + {\text H.c.} +  {\rm v} k  \sum_{\nu} s_{\nu} \eta^{\dagger}_{\nu,k} \sigma^z \eta_{\nu,k}. 
\end{equation}

For the case of the E$_u$ representation we have $\rm{v}=0$ and for the other representations we have $\rm{v}= \pm \Delta$. 

Defining the spinor $\eta_k=(\eta_{l,k,\uparrow}, \eta_{l,k,\downarrow},\eta_{r,k,\uparrow}, \eta_{ r,k,\downarrow})^T$, 
this effective Hamiltonian  can be expressed as
\begin{equation}\label{hkeff-tt-p-1}
  H^{p-p }_{ {\rm eff}, k} = \eta^{\dagger}_{k} \left[ t_{\rm J} \cos (\phi/2) \;  \tilde{\tau}^x \;  + 
  {\rm v} k   \eta^{\dagger}_{k} \; \tilde{\tau}^z \sigma^z \; \right] \eta_{k},
\end{equation}
where the Pauli matrices $\tilde{\tau}^j$ act on the left-right degrees of freedom. 
We see that it has the structure of a Dirac Hamiltonian with a mass term 
$\propto \cos (\phi/2)$.
The Hamiltonian of (\ref{hkeff-tt-p-1})  can be
diagonalized and has the 
following eigenenergies $\pm \varepsilon^{p-p}_{k}(\phi)$ with
\begin{equation}\label{andreevp}
\varepsilon^{p-p}_{k}(\phi)=  \sqrt{\left( {\rm v} k \right)^2 + t_{\rm J}^2  \cos^2(\phi/2)},
\end{equation}
which defines the Andreev spectrum. As a consequence of the dependence of the mass term with $\phi$ a gap opens  for arbitrary small $\phi$.
The derivative $\partial E_{k,-}/\partial \phi$ leads to a behavior of $J_k$ that is perfectly consistent with the behavior reported in Fig. \ref{figpk}. 
For the $E_u$ representation, the different amplitude of the discontinuity for different $k$-values can be explained by taking into account the renormalization of $t_{\rm J}$
due to the $k$-dependent projection of the edge-modes on the fermionic operators of the boundary in Eq. (\ref{ceffp}).

\subsubsection{ZKM model}\label{sec:tt-zkm}
\begin{figure}[h]
\begin{center}
\includegraphics[width=\columnwidth]{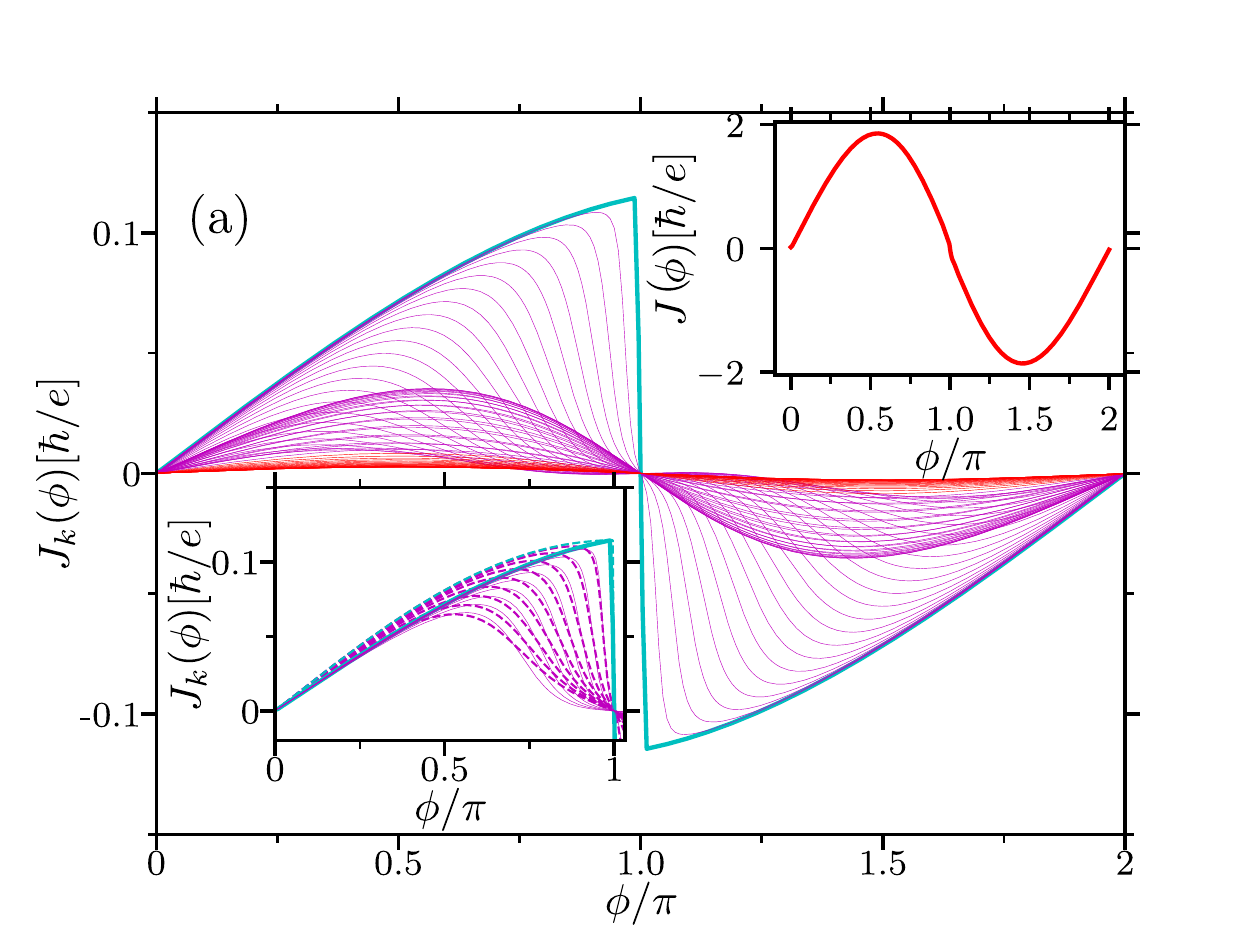}
\includegraphics[width=\columnwidth]{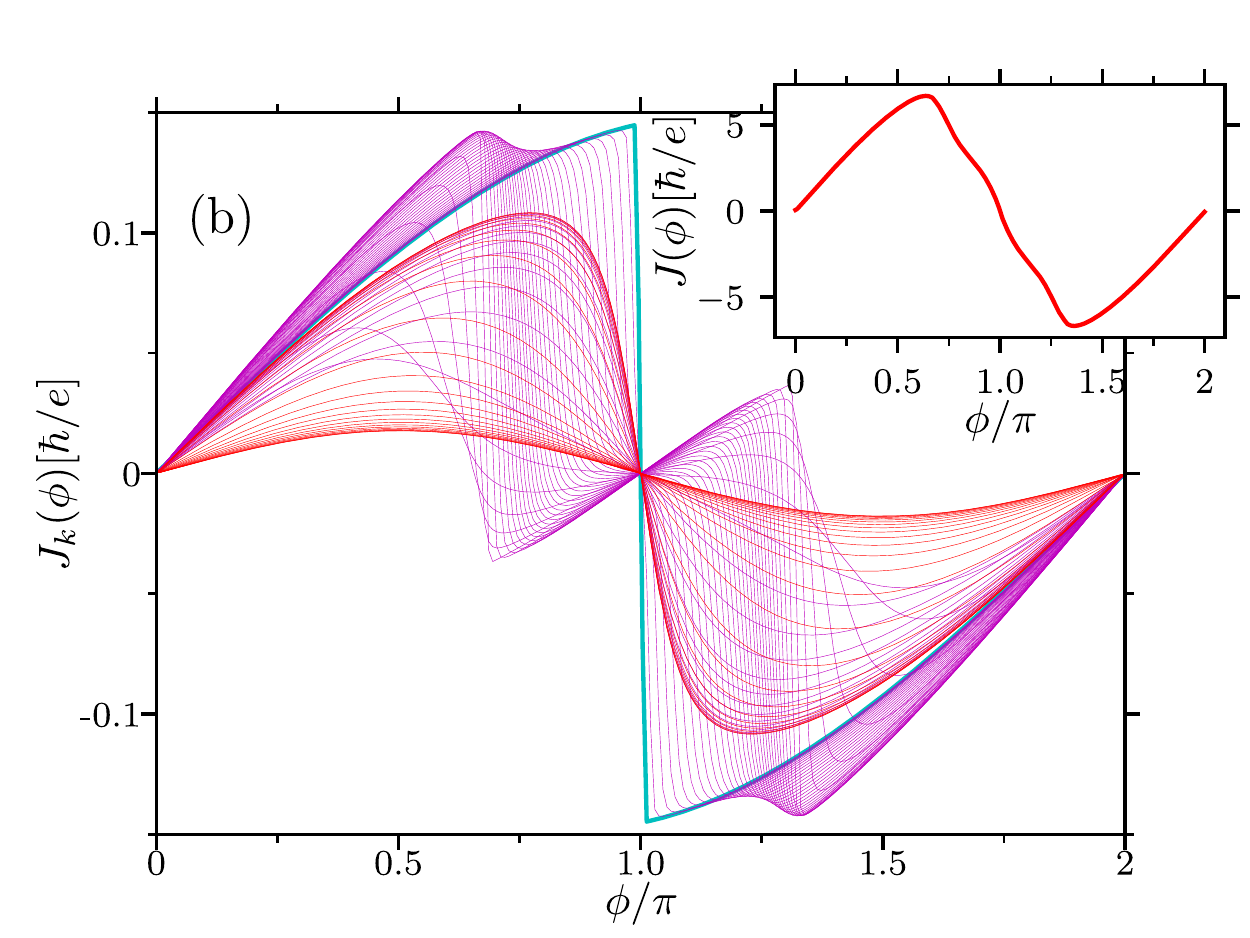}
\end{center}
\caption{(a) $k$-resolved Josephson current as a function of the phase difference in the topological phase for a junction of 2D topological superconductors with $t_J=t/2$,
$\Delta_0= 2\Delta_1$ and $\mu=\varepsilon_0$.
 The plot in thick lines corresponds to $k=\pi$. 
 Plots in violet correspond to the edge states, while the other $k$ values are shown in red. The upper insets show the total Josephson current and the effective Josephson current. 
 The lower inset  shows the comparison of the exact numerical solution with the prediction based on Eq. (\ref{Jeff}) with the effective parameters calculated with the exact solution as explained in Appendix \ref{ex-sol} (dashed lines). 
 (b) Same as top panel for $\Delta_0=\Delta_1$ and
$t_{\rm J}=t$. }
\label{fig2}
\end{figure}
The Josephson current for the different $k$-values as a function of the phase bias $\phi$ for the ZKM model is shown in Fig. \ref{fig2}. As in the previous section, we distinguish with different colors the contribution of the continuum states (red) and the edge modes (violet), highlighting the component of the zero modes corersponding to the time-reversal symmetric points $k_0=0,\pi$ (light blue). 
The latter mode presents the same type of discontinuity at $\phi=\pi$ observed in the $p$-wave models. We also observe   the typical $\propto \sin(\phi)$-behavior in the contribution of the states well inside the continuum. The behavior of the edge modes is more clearly distinguished for the parameters corresponding to the the upper panel and we will provide an analytical description below. The lower panel corresponds to parameters, for which the superconducting gap is smaller. In this case, there is a strong hybridization between the topological edge-states and those belonging to the quasiparticle continuum. We see interesting features, including several sign changes  of $J_k(\phi)$ for such mixed states. The total CPR for different parameters is shown in the top insets of both panels.

In order to analyze the contributions of the edge states in the present case, we follow the same  procedure of the previous section. 
Introducing the gauge transformation $\eta^{\dagger}_{\nu,k,s} \rightarrow e^{-i s s_{\nu} \pi/4}\eta^{\dagger}_{\nu,k,s}$ we have
\begin{equation}\label{fnuktil}
\tilde{f}_{\nu,k,s}=\frac{1}{\sqrt{2}}\eta_{\nu,k,s}, \;\;\;\; \tilde{f}^{\dagger}_{\nu,-k,s}
=-i s s_{\nu}\frac{1}{\sqrt{2}}\eta_{\nu,k,s},
\end{equation}
with the fermionic operators defined in Eq. (\ref{fermions}).


Therefore, assuming $\lambda \Delta_1 >0$,
\begin{eqnarray}\label{fnuk}
f_{\nu,k,\uparrow}&=&  \frac{e^{i  s_{\nu} \varphi_k/2}}{2}\left(\eta_{\nu,k,+}-
\eta_{\nu,k,-}\right),\nonumber \\
f_{\nu,k,\downarrow}&=& \frac{ e^{-i  s_{\nu} \varphi_k/2}}{2}\left(\eta_{\nu,k,+}+
\eta_{\nu,k,-}\right),\nonumber \\
f^{\dagger}_{\nu,-k,\uparrow}&=& -i s_{\nu}  \frac{e^{-i  s_{\nu} \varphi_k/2}}{2}\left(\eta_{\nu,k,+}+
\eta_{\nu,k,-}\right),\nonumber \\
f^{\dagger}_{\nu,-k,\downarrow}&=& -i s_{\nu} \frac{ e^{i  s_{\nu} \varphi_k/2}}{2}\left(\eta_{\nu,k,+}-
\eta_{\nu,k,-}\right).
\end{eqnarray}
Finally, we  use the relation to the parameters of Eq. (\ref{fnu}) corresponding to 
 the wave function of the lattice Hamiltonian, which leads to
\begin{eqnarray} \label{cend}
c^{\dagger}_{\nu,\pm k,\sigma}=\Omega_{\nu,k,\sigma}f^{\dagger}_{\nu,\pm k,\sigma}, \nonumber \\
\end{eqnarray}
with
\begin{equation} \label{wnu}
  w_{l,k}= \Omega_{l,k,\uparrow}= \mathcal{N}_{k}\sum_{\ell=1}^{2}\alpha _{k,\ell }=
   \overline{\Omega}_{r,k,\uparrow}= \overline{\Omega}_{l,k,\downarrow}=
   \Omega_{r,k,\downarrow}=\overline{w}_{r,k}.
\end{equation}
Substituting in  Eq. (\ref{junction}), and assuming that the right edge of S1 is connected to the left edge of S2, 
leads to the effective Hamiltonian for the junction. Including the contribution of the free edge states described by Eq. (\ref{edge-latt-1}) we get
\begin{eqnarray}\label{hkeff-tt}
   H_{{\rm eff}, k}^{\mathrm{ZKM-ZKM}} & = &  \cos(\phi/2) \sum_{s=\pm}  \left[t_{\mathrm{1},k} \eta^{\dagger}_{r,k,s}\eta_{l,k,s} 
   + i t_{\mathrm{2},k} \eta^{\dagger}_{r,k,s}\eta_{l,k,-s} + H.c.\right] \nonumber \\
   & + &  \sum_{s=\pm} s \varepsilon_{\lambda,k} ~\eta_{\nu,k,s}^{\dagger} \eta_{\nu,k,s},
\end{eqnarray}
where $\varepsilon_{\lambda,k}$ is defined in Eq. (\ref{ek}) and we have introduced the definitions
\begin{equation}
t_{\mathrm{1},k}= t_{\rm J} \mathrm{Re}(w_{k}^{2}e^{i s_r \varphi_k}),
~~~~~t_{\mathrm{2},k}= t_{\rm J} \mathrm{Im}(w_{k}^{2}e^{i s_r \varphi_k}).
\end{equation}
The term in the first line of Eq. (\ref{hkeff-tt}) describes the hybridization of the edge states through the Josephson-tunneling process, while the second one corresponds to the free edge states. 
In analogy to  the case of the $p$-wave model, we can introduce the spinor 
$\eta_k=(\eta_{l,k,+}, \eta_{l,k,-},\eta_{r,k,+}, \eta_{ r,k,-})^T$, in terms of which the effective Hamiltonian reads
\begin{equation}\label{hkeff-tt-1}
   H_{{\rm eff}, k}^{\mathrm{ZKM-ZKM}}  =  \eta^{\dagger}_{k} \; \left[\cos(\phi/2) \left(
  t_{\mathrm{1},k} \; \tilde{\tau}^x 
   + t_{\mathrm{2},k}  \; \tilde{\tau}^y \sigma^x \right) 
   +\; \varepsilon_{\lambda,k}  \; \sigma^z \;\right] \eta_{k}.
\end{equation}
We see that in the present case, the effective Hamiltonian for the coupled edge modes  has the structure of the Dirac Hamiltonian as in the 
case of the $p$-wave model, but with two mass terms. Both mass terms are $\propto \cos(\phi/2)$, which implies the opening of a gap in the Andreev spectrum for arbitrary small $\phi$. It is interesting to notice that, unlike the $p$-wave case, the  two massive terms are $k$-dependent 
in this case. This is a consequence of the 
 spin structure of the edge modes, which do not have a fixed direction in space, but have a $k$-dependent tilt $\varphi_k$.
The effective Hamiltonian can be diagonalized 
and has the following 
eigenenergies $\pm \varepsilon^{\rm ZKM}_{k,\pm }(\phi )$ with 
\begin{equation}
\varepsilon_{k,\pm }^{\rm ZKM} (\phi )=\sqrt{\left( t_{\mathrm{1},k}\cos (\phi /2)\pm
\varepsilon_{\lambda,k}\right) ^{2}+t_{\mathrm{2},k}^{2}\cos ^{2}(\phi /2)}.
\end{equation}%

The calculation of the Josephson current for this effective Hamiltonian results
\begin{eqnarray} \label{Jeff}
& & J_{{\rm eff},k}(\phi)=\frac{1}{2}t_{\rm eff}(\phi)\sin(\frac{\phi}{2}), \\
& & t_{\rm eff}(\phi)= \frac{\left[t_{{\rm 1},k} \cos(\phi/2) + \varepsilon_{\lambda,k} \right]t_{{\rm 1},k}+ t_{{\rm 2},k}^2  \cos(\phi/2)  }{\varepsilon^{\rm ZKM}_{k,+}(\phi)} \nonumber \\
& &~~~~~~~+\frac{\left[t_{{\rm 1},k} \cos(\phi/2) - \varepsilon_{\lambda,k} \right]t_{{\rm 1},k}+ t_{{\rm 2},k}^2  \cos(\phi/2)  }{\varepsilon^{\rm ZKM}_{k,-}(\phi)}.\nonumber
\end{eqnarray}

For the time-reversal symmetric points $k_0=0,\pm \pi$,  there is a level crossing in the spectrum because of which the ground state energy $E_{0, \rm eff}(k,\phi)$ has a cusp and its derivative 
is discontinuous at $\phi=\pi$, which explains the jump  in the Josephson current at this value of the phase. Other $k$-values corresponding to the edge modes are semi-quantitatively described by Eq. (\ref{Jeff}). 
An illustration is shown in the lower inset of  Fig. \ref{fig2} (a), where the Josephson current calculated from exact diagonalization of the full lattice model is explicitly compared with the prediction of Eq. (\ref{Jeff}) based on the analytical calculation of the parameters $w_k$ and $\varphi_k$ from Eqs. (\ref{norm}) and (\ref{rhodel}). Although these parameters depend in $k$, close to the Dirac point, such dependence can be neglected.
 We see that the agreement is very good and the slight quantitative mismatching can be understood by recalling that the analytical calculation introduces some approximations, namely it treats $\lambda$ perturbatively and also assumes strongly localized edge modes [see Eqs. (\ref{comm2}) and Eq. (\ref{rhodel})]. 
 The plots of Fig. \ref{fig2} (b) correspond to parameters for which the superconducting gap is smaller. Under these conditions, the topological edge 
 modes of each topological superconductor hybridize in the junction, not only with the topological edge states of the other superconductor but also with the 
 non-topological states above the gap. As a  consequence of this mixed hybridization 
other features, like sign changes and a saw-tooth type behavior observed in these plots emerge. This peculiar behavior can be qualitatively explained in terms of an effective Hamiltonian for the junction, which consists in adding a term representing the high-energy states
to the effective low-energy Hamiltonian of Eq. (\ref{hkeff-tt-1}). Such a procedure is similar  
 to the one explained in the next section for the description of the TRITOPS-S junction.

\begin{figure}[h]
\vspace{1.cm}
\begin{center}
\includegraphics[width=\columnwidth]{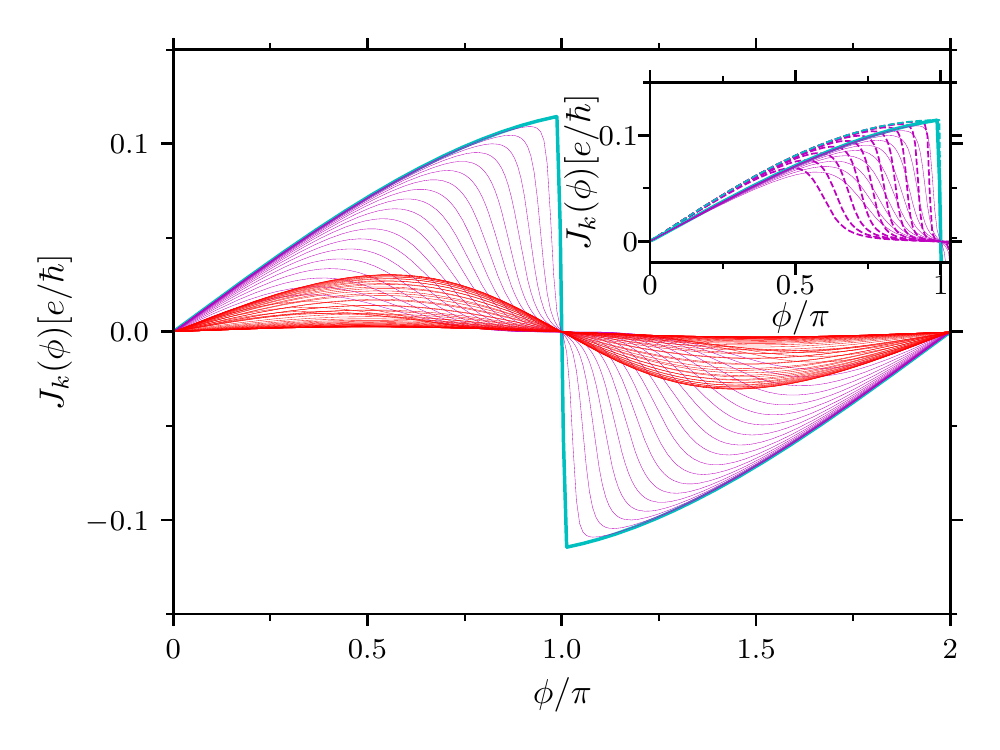}
\end{center}
\caption{$k$-resolved Josephson current for a junction tilted an angle $\beta=\pi/4$ respect to the xz-plane with $t_J=t/2$, $\Delta_0=2\Delta_1$, $\mu=\epsilon_0$. Inset: Comparison with the effective model. Dashed lines correspond to Eq. (\ref{Jeff}) with the parameters defined in Eq. (\ref{tilted}). }
\label{fig_tilted}
\end{figure}

We have considered so far junctions between TRITOPS with SOC oriented in the same direction. It is also interesting to consider a configuration where the two planes hosting the superconductors are tilted in an angle  $\beta$ around the $z$-axis in the coordinate frame  of Fig. \ref{fig:2} (b). Introducing such a rotation in the Hamiltonian of S2 in $H_k$ and in Eq. (\ref{junction}) leads to the $k$-resolved Josephson current shown in Fig. \ref{fig_tilted}. We appreciate
some different features for the $k$-values corresponding to the edge modes, in comparison to Fig. \ref{fig2} (a), which has been calculated for the same parameters of the Hamiltonian in a junction without any tilt ($\beta=0$).
As before, the behavior of $J_k(\phi)$ for $k$ belonging to the edge modes, can be captured with a good degree of accuracy by the description provided by the effective Hamiltonian describing the Josephson-tunnel coupled edge modes. In the present case, this corresponds to
Eq. (\ref{hkeff-tt}) suitably modified to account for the tilt, which implies modifying 
the parameters to
\begin{equation}\label{tilted}
t_{\mathrm{1},k}=t_{\mathrm{J}}\mathrm{Re}(w_{k}^{2}e^{-i\left(\varphi_k+\frac{\beta}{2}\right)}),
~~~~~t_{\mathrm{2},k}=t_{\mathrm{J}}\mathrm{Im}(w_{k}^{2}e^{-i\left(\varphi_k+\frac{\beta}{2}\right)}).
\end{equation}
This merely adds a shift $\beta/2$ to the tilt of the spins of the edge modes with respect to the plane of the superconductor. The corresponding
contribution to the
Josephson current  calculated from this effective model is given by Eq. (\ref{Jeff}) with these modified parameters. 


The  net Josephson current, resulting from adding the contribution of all the transverse $k$ channels is shown in the upper right inset of Fig. 5 (a) and (b). In the topological case, it shows a smooth but richer structure, which should be traced back to the maxima, minima and crossings that take place for the $k$ values corresponding to the edge states for these parameters. 

\subsection{TRITOPS-S junction} 
\begin{figure}[h]
\begin{center}
\includegraphics[width=\columnwidth]{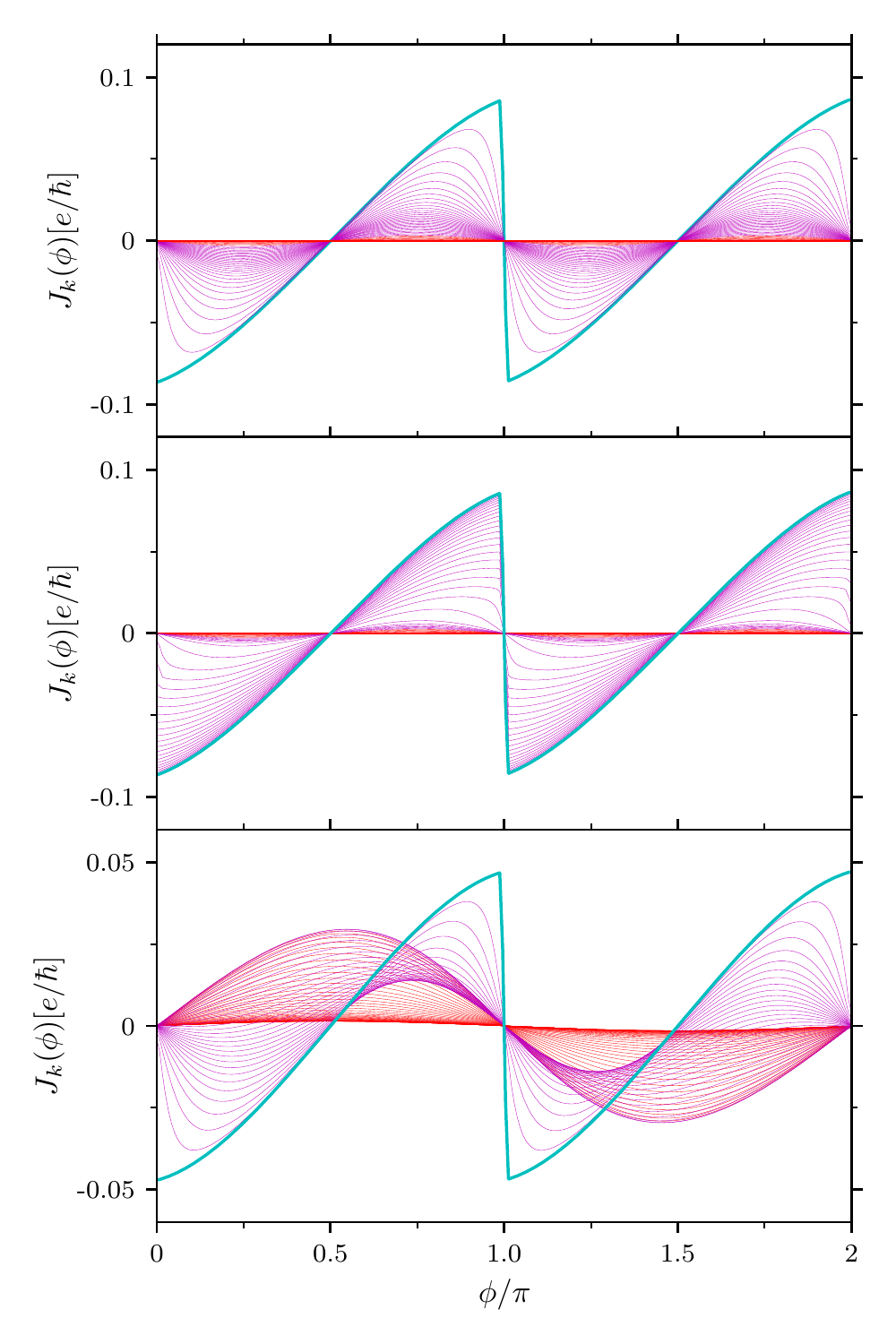}
\end{center}
\caption{$k$-resolved Josephson current as a function of the phase difference in the topological phase for a junction between a 2D topological superconductor with $p$-wave pairing (upper and middle panels) and a non-topological one with $s$-wave pairing.  These panels 
correspond to the representation $A_{1u}$ and $E_u$, respectively. The lower panel corresponds to a junction between the ZKM model and an ordinary superconductor (corresponding to $\Delta_1=\lambda=0$). The parameters are $t_J=\lambda=0.5t$, $\Delta=\Delta_0=2\Delta_1=0.4t$, $\mu=\Delta_0/\Delta_1 t$.}
\label{figkts}
\end{figure}

We now consider a Josephson junction between a TRITOPS and a non-topological superconductor. Concretely, we consider the Hamiltonian  of
Eqs. (\ref{hkfull}) and (\ref{junction}) with S1 being a BCS superconductor with only local pairing $\Delta_0$. This corresponds to
Eq. (\ref{h-zkm}) with $\lambda=\Delta_1=0$. The results for the $k$-resolved Josephson current for S1 modeled by the three TRITOPS Hamiltonians 
studied in the previous sections are shown in Fig. \ref{figkts}.

\begin{figure}[h]
\begin{center}
\includegraphics[width=\columnwidth]{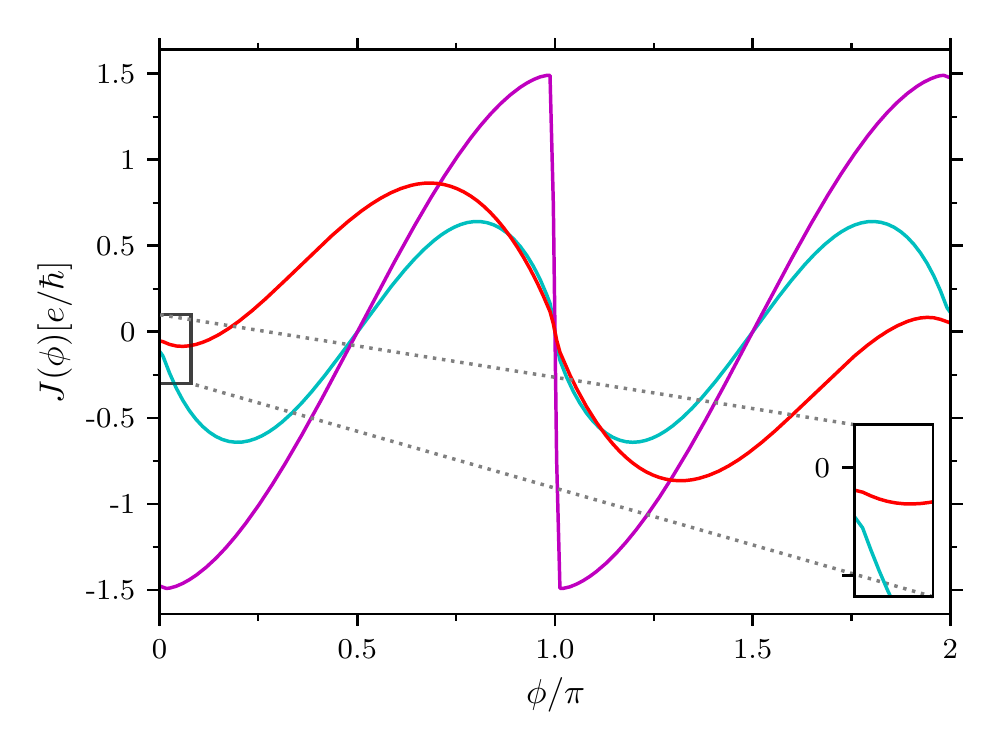}
\end{center}
\caption{Total Josephson current as a function of the phase difference in the topological phase for a junction between a 2D topological superconductor and a non-topological one with $s$-wave pairing. Light blue, violet and red colors correspond to the $A_{1u}$, $E_u$ and $ZKM$ models, respectively.}
\label{figktst}
\end{figure}


The common pattern we can identify in these three configurations is the behavior of the $k_0$-component. It is characterized by three remarkable features, 
in striking contrast with  the TRITOPS-TRITOPS junctions.
These are: (i) a sign change in $J_{k_0}(\phi)$, (ii) twice the periodicity in $\phi$ and (iii) discontinuous jump at $\phi=0$. 
Such a peculiar behavior was discussed in the context of 1D systems in Refs. \cite{qi2009time,chung2013time,haim2019time}. 
In the case of the $E_u$ representation, these features are observed, not only for $k_0$, but also for all the $k$-values belonging to the (zero-energy) edge modes.
A similar behavior was discussed  in 
the framework of 
Josephson junctions between a 
superconductor with d$_{x^2-y^2}$ pairing and superconductors with $s$-wave pairing \cite{tanaka1,tanaka2,tanaka3}. A jump in the CPR akin to the one observed in Fig. \ref{figkts} is predicted
when the nodal line of the  $d$-wave superconductor is perpendicular to the junction, as a consequence of zero modes in the interface. In our case, the existence of the zero modes are associated to
the topological edge states.
 It is important to notice that this feature is, however, different from the so called "anomalous Josephson effect" taking place when time-reversal symmetry is broken in the superconductor and/or in the junction at $\phi=0$ \cite{alidoust2018strain,alidoust2020critical,alidoust2021cubic,zazunov2009anomalous}. In fact, as a consequence of the time-reversal symmetry,
the CPR in the TRITOPS-S junction obeys $J(\phi=0)=0, \forall k$. A finite, albeit arbitrary small $\phi$ is necessary to induce the jump in $J_{k}(\phi)$ for $k$ belonging to the zero-energy modes.


We can also derive an effective low-energy Hamiltonian for the TRITOPS-S junction. To this end, we consider the edge modes of the topological side, coupled to the high-energy quasiparticle excitations of the non-topological (S) one. In order to simplify the calculations, we neglect the free dispersion relation and we consider the following model for the S-side
\begin{equation}\label{hsup}
H_{S,k}=\Delta _{0}(c_{k,\uparrow }^{\dagger }c_{-k,\downarrow }^{\dagger }+%
\mathrm{H.c.})=\sum_{s=\pm }s \Delta _{0} \; d_{k,s}^{\dagger } d_{k,s}+%
\mathrm{constant},
\end{equation}%
with $d_{k,s}=\left( c_{k,\uparrow }\pm c_{-k,\downarrow }^{\dagger
}\right) /\sqrt{2}$.


Considering the Hamiltonian of Eq. (\ref{gammanu}) for the edge states of the $p$-wave Hamiltonian expressed in the 
basis of the spinor $\eta_{\nu,k}=(\eta_{\nu,k,\uparrow}, \eta_{\nu,k,\downarrow})^T$
and integrating out the degrees of freedom of the ordinary superconductor in the second order of perturbation theory in the tunneling coupling $%
t_{\mathrm{J}}$, we get
\begin{equation}\label{p-s}
    H_k^{p-\mathrm{S,eff}}=\eta^{\dagger}_k \left[\tilde{\varepsilon}^p_{k} \sigma^z + m^p\sin (\phi ) \sigma^y \right] \eta_k
\end{equation}
with 
\begin{equation}\label{para-p}
  \tilde{\varepsilon}^p_{k}={\rm v} k  s_{\nu}, \;\;\;\;\; m^p=-|t_{\rm J}|^2/\Delta_0.
\end{equation}

For the case of the ZKM model, we can consider the projections on the edge modes of the fermionic operators at the end of the TRITOPS   by using 
Eqs. (\ref{fnuktil}), (\ref{fnuk}) and  (\ref{cend}) and integrate out the fermions of the
S-side in a similar way as before. This leads to the following effective Hamiltonian for the
junction expressed in the basis of the  spinor $\eta_k=( \eta_{\nu,k,+}, \eta_{\nu,k,-})^T$ (see Appendix \ref{appb} for details)
\begin{equation}\label{zkm-s}
H_{\nu ,k}^{\mathrm{ZKM-S,eff}}= \eta^{\dagger}_k \left[\tilde{\varepsilon}^{\rm ZKM} \sigma^z + m^{\rm ZKM}_{k}\sin (\phi ) \sigma^x \right] \eta_k.  
\end{equation}
where we adopt the same notation as in Sec. \ref{sec:tt-zkm} and we have introduced 
\begin{equation}\label{para-zkm}
\tilde{\varepsilon}^{\rm ZKM}_{k}=\varepsilon_{\lambda,k}\left( 1-m^{\rm ZKM}/\Delta _{0}\right) \text{, }\;\;\;\; m^{\rm ZKM}=2s_{\nu
}t_{\mathrm{J}}^{2}|w_{k}|^{2}/\Delta _{0}.  
\end{equation}%

The diagonalization of the two Hamiltonians defined in Eqs. (\ref{p-s}) and (\ref{zkm-s}) for the TRITOPS-S junction leads to the eigenstates $\pm \varepsilon^{\rm T-S}_{k}(\phi )$
with 
\begin{equation}\label{epts}
\varepsilon^{\rm T-S}_{k}(\phi )=\sqrt{\left(m^{\rm ZKM}\right)^{2}\sin ^{2}(\phi )+\tilde{\varepsilon}_k^{2}},
\end{equation}
with the parameters defined in Eqs. (\ref{para-p}) and (\ref{para-zkm}), for the $p$-wave and ZKM Hamiltonians for the TRITOPS, respectively. This leads to the many-body ground-state energy 
$E_{0}(\phi ) \equiv -\varepsilon^{\rm T-S}_{k}(\phi )$. 
Hence, the Josephson CPR is given by 
\begin{equation}
J_{\mathrm{eff},k}(\phi )=-\frac{1}{2\varepsilon^{\rm T-S}_{k}(\phi )} \left(m^{\rm ZKM}\right)^{2}\sin(2\phi ). \label{Jeff-ts}
\end{equation}
This analytical expression is in full agreement with the behavior of the Josephson current for $k$ corresponding to the edge states shown in Fig. \ref{figkts}. In the case of the
$p$-wave pairing within the E$_u$ representation, which is shown in the middle panel, all the $k$-components close to the Dirac point have a jump at $\phi=0$ because the edge channels are dispersionless, hence $\tilde{\varepsilon}_k=0$ in Eq. (\ref{epts}).  

To finalize, we show in Fig. \ref{figktst} the total CPR obtained by adding all the $k$-components for the three models. We see that all the three cases are characterized by a jump at $\phi=0,\; \mbox{mod} (\pi)$. As expected, the amplitude of this feature is much more pronounced in the case of the E$_u$-type $p$-wave pairing.

\section{Instabilities and bulk-boundary correspondence}
In this section we discuss how our findings are related to the expectations that follow from the bulk-boundary correspondence. We will see that the behavior at TRITOPS-TRITOPS junctions is fully in line with the bulk-boundary correspondence. The junction separates two topologically non-trivial systems. Hence states in the junction area  should be gapped. The tunneling term between two TRITOPSs must therefore induce a gap of the edge states of both topological superconductors.  By the same logic one would expect massless states at the TRITOPS-S junction, as it  separates a topologically trivial and non-trivial state. However, our results imply that  TRITOPS-S junctions  violate the bulk boundary principle. We will argue that this is a consequence of the spontaneously broken time-reversal symmetry in the TRITOPS-S junction itself. With the protecting symmetry spontaneously broken at the edge, edge states become massive. 
We expect this to be a fully generic feature of TRITOPS-S junctions.

 The low energy states of both junctions can be described in terms of a  Majorana spinor with Hamiltonian
\begin{equation}
    H_{\rm edge}=vp \alpha+m\left(\phi\right)v^{2}\beta.
    \label{eq:Dirac_at_junction}
\end{equation}
For the TRITOPS-TRITOPS junction we have a four component Majorana spinor.
We focus on the $p$-wave case Eq. (40), for which $\alpha= \tilde{\tau}^z \sigma^z$ and $\beta=\tilde{\tau}^x$. In addition, the fermion mass depends on the relative phase according to $m\left(\phi\right)=m_{0}\cos\left(\phi/2\right)$, where $m_0$ is linearly proportional to the tunneling matrix element $t_J$.
In distinction, at the TRITOPS-S junction we have a two-component Majorana spinor 
with $\alpha=\sigma^{z}$, $\beta=\sigma^{x}$. The most crucial difference is  the phase dependence of the
mass $m\left(\phi\right)=m_{0}\sin\phi$, where $m_0\propto t_J^2/\Delta_0$ with pairing gap $\Delta_0$ of the topologically trivial superconductor.

So far, we have considered the phase bias of the junction as a parameter that is fixed by external conditions, like implementing the junction in a ring-shape structure threaded by a magnetic flux.  We now consider the 
 junction regarding $\phi$ as an internal degree of freedom, in order to analyze the stability close to $\phi=0$. The usual phase dependence of the Josephson energy is
\begin{equation}
E_{J,0}\left(\phi\right)=\frac{1}{2\lambda_{J}^{2}}\left(1-\cos\phi\right),
\label{eq:junction_bare_energy}
\end{equation}
which yields an equilibrium phase $\phi^{*}=0\,{\rm mod}\left(2\pi\right)$.
Eq.\eqref{eq:junction_bare_energy} is the result of tunneling due to states above the bulk pairing gap, just like in any other  superconductor.
Using this value for the equilibrium phase difference $\phi^*$ and inserting in the two mass-phase relations for the two junction types, edge states   are massive for TRITOPS-TRITOPS junctions ($\cos(\phi^*/2)\neq 0$) and massless  for TRITOPS-S junctions ($\sin(\phi^*)= 0$). This is in line with the
expectation that follows from the bulk-boundary correspondence.

It is however important to analyze the impact of the edge states on the current-phase relation, i.e. to properly include the edge-state contribution to the Josephson energy:
\begin{equation}
E_{J}\left(\phi\right)=E_{J,0}\left(\phi\right)+\delta E_{J}\left(\phi\right).
\end{equation}
Here,
\begin{equation}
  \delta E_{J}\left(\phi\right)=\langle H_{\rm edge} \rangle_\phi-\langle H_{\rm edge} \rangle_{\phi=0}  
\end{equation}
is the phase-dependent expectation value of the energy due to edge modes. We made the assumption that the phase stays constant in space  along the junction, an assumption that we will relax below.

In order to determine $\delta E_{J}\left(\phi\right)$ we integrate out the edge state
fermions. To justify this we assume and check later for consistency that the fermion mass is finite. It follows from  $H_{\rm edge}$ of Eq.\eqref{eq:Dirac_at_junction} that
\begin{equation}
\delta E_{J}\left(\phi\right)=-\frac{v}{8\pi}m\left(\phi\right)^{2}
\log\left(\frac{\Lambda^{2}}{m\left(\phi\right)^2}\right).
\label{eq:Jintcorre}
\end{equation}
Here $\Lambda$ is the high-energy cutoff.

Let us first comment on the impact of edge modes on the current-phase relation of TRITOPS-TRITOPS junctions. If we use $m\left(\phi\right)=m_{0}\cos\left(\phi/2\right)$ in Eq.\eqref{eq:Jintcorre}, the minimum in $E_J\left(\phi\right)$ continues to be at $\phi=0 \,{\rm mod}(2\pi)$ and the edge modes are indeed massive $m\left(\phi^*\right)=m_0\neq 0$, fully consistent with the bulk-boundary correspondence.
Only near $\phi=\pi$, where the fermion mass changes sign,  do we find  a singular behavior for the current:
\begin{equation}
 J\left(\phi \right) \sim -\log\left(\frac{\Lambda^2}{m_0^2(\phi-\pi)^2}\right)\left(\phi-\pi\right)
\end{equation}
This is the main effect of edge modes for TRITOPS-TRITOPS junctions.

More dramatic behavior occurs at the TRITOPS-S junction.
With $m\left(\phi\right)=m_{0}\sin\phi$, one easily finds  that the  singular logarithmic dependence near $\phi=0$
implies that  the minimum in $E_{J}\left(\phi\right)$ is always shifted to a finite phase $\phi^{*}$, yielding a finite fermion mass, which corresponds to broken chiral symmetry. Using $\cal{C}=\cal{P}\cal{T}$ for the chiral, parity, and time-reversal symmetries, we see that broken $\cal{C}$ with intact $\cal{P}$, breaks the  time-reversal symmetry, as expected for a junction with non-trivial  phase difference $\phi\neq0,\pi$. Clearly  the bulk-boundary correspondence does not apply.

We conclude that the edge modes at the junction between a TRITOPS and a conventional superconductor are not gapless, but massive which is closely connected to a finite equilibrium phase difference at the junction. The bulk boundary corresponence at the edge is invalidated as the gapless modes are unstable against an infinitesimal Josephson coupling. The protecting time-reversal symmetry is broken at the junction as a consequence of the phase-edge mode coupling. 

Other physical effects may play a role. Particularly interesting are the role of phase fluctuations, which could induce mechanisms tending to restore the broken time-reversal symmetry in the TRITOPS-S junction.  Another  interesting effect that could take place in the coupled dynamics of the edge states and the phase fluctuations is the emergence of Majorana zero modes that are tied to solitonic phase slips. Those localized zero modes are expected for both junction types.  All these phenomena are worth of being analyzed in combination with capacitive 
electron-electron interactions in the junction. The starting point to this goal are the effective Hamiltonians of Eqs. 
(\ref{hkeff-tt-p-1}), (\ref{hkeff-tt-1}), (\ref{p-s}) and (\ref{zkm-s})
 we have derived for the different junctions suitable extended to address these other effects.

\section{Conclusions}
We have studied different models for two dimensional time-reversal symmetric topological superconductors (TRITOPS), with and without spin-orbit coupling.
To this end,  we have derived effective Hamiltonians for the edge modes and shown that their spectrum and spin texture strongly depend on the point-group symmetry of the superconducting pairing as well as on the spin-orbit coupling.\\
We have then analyzed
wide Josephson junctions between two topological superconductors (TRITOPS-TRITOPS) as well between a topological 
superconductor and an ordinary s-wave superconductor (TRITOPS-S).
The discontinuous current-phase relation near $\phi=\pi$, known from junctions between two one-dimensional topological superconductors, continues to be visible in our wide TRITOPS-TRITOPS junctions. It is particularly pronounced in the two-component, nematic superconductors, where edge modes are non-dispersive. This result follows not only from our approximate continuum's theory, but is equally present in the full numerical solution of the lattice version. 
Singular behavior near  $\phi=\pi$ is however also present in other junctions between two TRITOPSs, albeit weaker, given the edge-state dispersion. This behavior is caused by vanishing mass of the edge modes at $\phi=\pi$.\\
The behavior is rather different in TRITOPS-S junctions between a time-reversal symmetric topological  and a topologically trivial superconductor. Now 
the edge-mode gap vanishes for $\phi=0$, and minimization of the energy leads to a finite but small $\phi$ leading to time-reversal symmetry breaking in the edge. This endows the edge modes with a mass and generates an unexpected jump in the Josephson current.

These results provide useful hints in the experimental search of the TRITOPS phase. In addition,
the effective Hamiltonians for the different junctions
we have derived in the present work  are the foundation stones to investigate several other interesting phenomena that may take place in Josephson junctions with time-reversal symmetric topological superconductors, and can be extended to analyze the effect of phase fluctuations, the generation of solitonic modes and charging effects.


\section*{Acknowledgments}
We are grateful to Dmitriy S. Shapiro and Alexander Shnirman for helpful discussions. We acknowledge financial support provided by PICT 2017-2726 and PICT 2018-01546 of the FonCyT, and CONICET Argentina. We also thank the support 
of SPUK collaboration (JS and LA) and the Alexander von Humboldt Foundation (LA).

\appendix
\section{Details of the derivation of the edge states of the $p$-wave model}\label{papp}
We assume $\Delta >0$, and the topological phase corresponds to $\mu >0$. We focus on an edge intersecting the horizontal axis at the $x=0$ of a slab of infinite length along the $y$-direction. 
To analyze the right/left edge, we consider
 a domain wall of the form $\mu^{r/l}(x)= \mp \mbox{sgn}(x)|\mu_0|$, which corresponds to the topological phase in the region with $x<0~/ ~x>0$, respectively. 
 
 For  $p_y=0$, there exists a Kramer's pair of Majorana zero modes, which can be calculated from the solution of
\begin{equation}\label{hp0}
\left\{ -\mu^{\nu}(x) \tau^z \sigma^0 + \Delta \tau^x \left(-i \partial_x \sigma^x  \right) \right\} \Phi^{\nu}_0(x)=0,
\end{equation}
where we have neglected, for simplicity, the dispersion relation. 
The solutions are
\begin{equation}\label{zero-p-z}
\Phi_{0,s}^{\nu}(x)= g^{\nu}(x) \Lambda^{\nu}_{0 s}, ~~~~~~g^{\nu}(x)=g_0 e^{ s_{\nu} \int_0^{x} dx^{\prime} \frac{\mu^{\nu}(x^{\prime})}{\Delta}}.
\end{equation}
with $\Lambda^{\nu}_{0, s}$ being a spinor that satisfies
$\tau^y \sigma^x \Lambda^{\nu}_{0 s}=s_{\nu} \Lambda^{\nu}_{0 s}$, with $s_r=-s_l=1$ and $s=\pm$. Hence, $\Lambda^{\nu}_{0 s}=\frac{1}{2}\left(1,s,s_{\nu}si,s_{\nu}i \right)^T$.
For finite $p_y$ we look for solutions of the form

\begin{equation}\label{field}
    \Psi({\bf x},t)= \sum_{p_y >0} e^{-i E^\nu_{p_y} t} \left(\Phi^\nu_{p_y}({\bf x})\eta_{\nu, p_y} + C{\cal T}\Phi^\nu_{p_y}({\bf x})\eta^\dagger_{\nu,p_y}\right),
\end{equation}

being $C{\cal T}\equiv -i\tau^y i \sigma^y {\cal K}$ where $C$ is the  charge conjugation, ${\cal T}$ is the time-reversal operator in Nambu space, ${\cal K}$ is complex conjugation and the Bogoliubov operator is $\eta_{\nu,p_y}=\int d^2x \Phi^{\nu\dagger}_{p_y}({\bf x})\Psi({\bf x})$.
Hence, we must solve
\begin{equation}\label{trans}
\left\{-\mu^{\nu}(x) \tau^z \sigma^0 +\Delta \tau^x \left(- i \partial_x \sigma^x \mp i \partial_y \sigma^y \right)\right\}\Phi^{\nu}_{p_y}(\mathbf{x})=E^{\nu}_{p_y} \Phi^{\nu}_{p_y}(\mathbf{x}).
\end{equation}
We find two degenerate solutions, which we label with $\sigma=\uparrow, \downarrow$ for each $p_y$. 
The eigenenergies are
\begin{equation}
E^{\nu}_{p_y, \sigma}=\pm {\rm v}_{\nu,\sigma} p_y
\end{equation}
with ${\rm v}_{\nu,\sigma}= s_{\nu} s_{\sigma} \Delta$, with $s_{\uparrow}=-s_{\downarrow}=1$. The eigenfunctions are
\begin{equation}
\Phi^{\nu}_{p_y, \sigma}(x,y)=g^{\nu}(x) e^{is_\nu s_\sigma \pi/4} e^{i p_y y} \Lambda^{\nu}_{0,\sigma},~~~~\Lambda_{0,\sigma}^{\nu}=\frac{\left(\Lambda_{0,+}^{\nu} +s_{\sigma} \Lambda_{0,-}^{\nu} \right)}{\sqrt{2}}.
\end{equation}
The corresponding Bogoliubov  operators are 
\begin{equation} \label{gammanuap}
\eta_{\nu,p_y,\sigma}=\frac{e^{is_\nu s_\sigma \pi/4}}{\sqrt{2}} \left(c_{\nu,p_y,\sigma}-i s_{\nu} s_\sigma c^{\dagger}_{\nu,-p_y,\sigma}\right).
\end{equation}

\section{Derivation of an approximate continuum Hamiltonian for the ZKM model}\label{zkm-cont}
We find it convenient to transform the Hamiltonian of  Eq. (\ref{h-zkm}) by means of a rotation $R=e^{-i \pi/4 \sigma^x}$ in the spin degrees of freedom, which transforms $\sigma^y \rightarrow \sigma^z$.
The derivation of the continuum Hamiltonian is particularly simple for
$\Delta_0=\pm 2\Delta_1$ and we shall focus on $\lambda,\; \Delta_1 >0$. Let us assume, for concreteness, the case with 
$\Delta_0=-2 \Delta_1, ~\lambda >0$,  and $k_y=0$, in which case the nodal surface crosses at the
nodal points $ (\pm \pi/2,0)$ and the two bands have a well defined $z$-component of the spin $\pm$ 1/2.

For $|\mu +2 t|\leq 2 \lambda$,  there are four  Fermi points in the $k_x$-axis (see Fig. \ref{fig:2}.a).
 We call them $\pm k^F_1, \pm k^F_2$, with $k^F_1$ and $k^F_2$
 belonging to the different branches $\uparrow, \downarrow$, respectively. Hence, linearizing with respect to the Fermi points 
 the
spectrum without pairing has right  and left movers with $\uparrow, \downarrow$ $z$-component of spin.
Projecting the pairing potential on the Fermi points and expanding  with respect to the points $\pm \pi/2$, which are precisely the nodal points of this potential, we have the following low-energy Hamiltonian 
\begin{equation}\label{hpx}
H^{k_y=0}_{\bf p}=  -\delta \mu \tau^z \sigma^0 + 2 \Delta_1 p_x \tau^x \sigma^z+ 2\lambda p_y \tau^z \sigma^x,
\end{equation}
which is defined in the basis of the spinor 
$(c_{k^F_1, \uparrow}, c_{k^F_2, \downarrow}, c_{-k^F_1, \downarrow}^\dagger, -c_{-k^F_2, \uparrow}^\dagger)^T$ 
with $|p_x|=k_1^F - \pi/2 =\pi/2-k_2^F$, $p_y=k_y$, ${\bf p}=(p_x,p_y)$ and 
$\delta \mu=\mu + 2t -2 \lambda$.

We can repeat the argument along  $k_y$ for $k_x=0$. The difference is that the two bands represented in
Fig. \ref{fig:2}.a have spin components along the $x$-direction instead of $z$.  The resulting Hamiltonian is
\begin{equation}\label{hpz}
H^{k_x=0}_{\bf p}=  \delta \mu \tau^z \sigma^0 + 2 \Delta_1 p_y \tau^x \sigma^x-2\lambda p_x \tau^z \sigma^z ,
\end{equation}
with $p_y=k_y \mp \pi/2$, $p_x=k_x$. 
Furthermore, a similar reasoning can be followed for the case with $\Delta_0 = 2 \Delta_1$ for $|\mu-2t| \leq 2\lambda$ and the same values of the other parameters. 
 For simplicity, we have neglected the momentum dependence of the dispersion relation of the two bands without pairing at the Fermi energy.

 We now consider the Hamiltonian $H^{k_y=0}_{\bf p}$ given by (Eq. \ref{hpx}) to derive the wave function and the effective Hamiltonian for the edge states along the $y$-direction. 
 The structure of the solution for the case  $p_y=0$ is identical to Eq. (\ref{hp0}). In turn, as pointed out in the previous section, this  solution has an identical structure as Eq. (\ref{zero-p-z}), but
 $\Lambda^{\nu}_{0, \sigma}$ is now a spinor that satisfies
$\tau^y \sigma^z \Lambda^{\nu}_{0 \sigma}=s_{\nu} \Lambda^{\nu}_{0 \sigma}$, with $s_r=-s_l=1$ and $\sigma=\uparrow, \downarrow$. Hence, $\Lambda^{\nu}_{0 \uparrow}=\left(1,0,s_{\nu}i,0 \right)$
and $\Lambda^{\nu}_{0 \downarrow}=\left(0,1,0,-s_{\nu}i \right)$. For $p_y\neq 0$, we consider a field operator with the structure of Eq. (\ref{field}), where the spinors should satisfy
\begin{equation}\label{trans1}
\left\{-\mu^{\nu}(x) \tau^z \sigma^0 +2\Delta_1 \tau^x \left(-i \partial_x \sigma^z\right)+ 2\lambda p_y \tau^z \sigma^x \right\}\Phi^{\nu}_{p_y}(x)=E^{\nu}_{p_y} \Phi^{\nu}_{p_y}(x).
\end{equation}
We find two degenerate solutions, which we label with $s=+,-$ for each $p_y$. 
The eigenenergies are
\begin{equation}
E^{\nu}_{p_y, s}={\rm v}_{s} p_y
\end{equation}
with ${\rm v}_{s}=s 2 \lambda$. The eigenfunctions are
\begin{equation}
\Phi^{\nu}_{p_y, s}(x,z)=g^{\nu}(x) e^{-i s s_{\nu} \pi/4} e^{i p_y y} \Lambda^{\nu}_{0,s},~~~~\Lambda_{0,s}^{\nu}=\frac{\left(\Lambda_{0,\uparrow}^{\nu} 
+s\Lambda_{0,\downarrow}^{\nu} \right)}{\sqrt{2}}
\end{equation}
and the corresponding Bogoliubov  operators can be expressed as
\begin{eqnarray} \label{gammanux}
\eta_{\nu,p_y,s}&=& \frac{e^{-is s_{\nu} \pi/4}}{\sqrt{2}} \left(c_{\nu,p_y,s}+i s s_{\nu}  c^{\dagger}_{\nu,-p_y,s}\right),\nonumber \\
c_{\nu,p_y,s}&=&\frac{1}{\sqrt{2}} \left(c_{\nu,p_y,\uparrow}+ s c_{\nu,p_y,\downarrow}\right),~~~~~\Delta_0/\Delta_1<0.
\end{eqnarray}
The corresponding effective Hamiltonians for the edges are given in Eq. (\ref{edge-zkm}).

  For the case with $\Delta_0/\Delta_1>0$, we can follow a similar approach, taking into account that  the expansion leading to the  effective continuum Hamiltonian must be done with respect to ${\bf k}_0=(\pi, \pi)$. In such a case, we would get for $H^{k_y=\pi}_{\bf p}$, with $p_x=k_x \pm \pi/2$, $p_y=k_y-\pi$ and $H^{k_x=\pi}_{\bf p}$,with $p_x=k_x - \pi$, $p_y=k_y \pm \pi/2$  expressions like those of Eqs. (\ref{hpx}) and (\ref{hpz}) but with opposite sign of the pairing term.

\section{Exact solution of the ZKM Hamiltonian for a  transverse channel}\label{ex-sol}
We consider the ZKM model in a ribbon of finite length $N_x$ along the $x$- direction and periodic boundary conditions in the transverse direction. For a single $k$-channel as defined in Eq. (\ref{h-zkm-1}) this Hamiltonian is one-dimensional. Hence, it is possible to solve it with a similar method as the one introduced in Ref. \cite{alaseprl,alaseprb}. The procedure is very similar as the one explained for this specific model in Ref. \cite{aligia2018entangled}. We explain bellow the main steps.

We express the Hamiltonian as follows
\begin{equation}
H_k=\sum_{j} H_{kj}^{a}+H_{kj}^{b}+H_{kj}^{\lambda },
\label{hs}
\end{equation}
where the first term is
\begin{eqnarray}
H_{kj}^{a} &=&\xi _{k}\sum\limits_{\sigma }c_{kj\sigma }^{\dagger
}c_{kj\sigma }+\left( \Delta _{k}c_{kj\uparrow }^{\dagger }c_{-kj\downarrow
}^{\dagger }+\text{H.c.}\right) ,  \label{xik}
\end{eqnarray}
the second one is
\begin{eqnarray}\label{h1}
H_{kj}^{b} &=&-t\left( c_{kj+1\sigma }^{\dagger }c_{kj\sigma }+\text{H.c.}%
\right) +\left[ -i\lambda \left( c_{kj+1\uparrow }^{\dagger }c_{kj\uparrow
}-c_{kj+1\downarrow }^{\dagger }c_{kj\downarrow }\right) \right.
 \nonumber  \\
&&\left. +\Delta _{1}\left( c_{kj\uparrow }^{\dagger }c_{-kj+1\downarrow
}^{\dagger }+c_{-kj+1\uparrow }^{\dagger }c_{kj\downarrow }^{\dagger }
\right)+\text{H.c.}\right]
\end{eqnarray}
These two terms are combined as follows
\begin{equation}
H_{k}^{0}=\sum\limits_{j}\left(
H_{kj}^{a}+H_{kj}^{b}+H_{-kj}^{a}+H_{-kj}^{b}\right),  \label{h0k}
\end{equation}
and this Hamiltonian is solved exactly.
The third term is 
\begin{equation}
H_{kj}^{\lambda }=-2\lambda \sin k\left( c_{kj\uparrow }^{\dagger
}c_{kj\downarrow }+c_{kj\downarrow }^{\dagger }c_{kj\uparrow }\right),
\label{hlam}
\end{equation}
and it is treated as a perturbation, by defining
\begin{equation}
H_{k}^{\lambda }=\sum\limits_{j}\left( H_{kj}^{\lambda }+H_{-kj}^{\lambda
}\right).  \label{hlk}
\end{equation}
As in previous works \cite{aligia2018entangled,aligia2020tomography}, the Hamiltonian matrix is expressed in a basis
constructed by mapping the
annihilation ($a$) and creation ($c$) operators to different states
\begin{equation}
c_{\alpha }\leftrightarrow |\alpha a\rangle \text{, }c_{\alpha }^{\dagger
}\leftrightarrow |\alpha c\rangle.  \label{map}
\end{equation}%
A solution for $H_{k}^{0}$ in a chain of $N_x$ sites with open boundary conditions with the structure of states of the generalized Bloch form
\begin{equation}
|zk\sigma b\rangle =\sum\limits_{j=1}^{N_{x}}z^{j-1}|kj\sigma b\rangle,
\label{bloch}
\end{equation}%
where $b=a$ or $c$, is proposed. Following the same steps as in Refs. \cite{aligia2018entangled,aligia2020tomography}
a zero mode localized at
the left ($L$) side of the stripe is obtained:
\begin{equation}
\gamma _{Lk\uparrow }=N_{k}\sum\limits_{i=1}^{2}\alpha
_{i}\sum\limits_{j=1}^{N_{x}}z_{i}^{j-1}\left( c_{kj\uparrow
}+i\zeta c_{-kj\downarrow }^{\dagger }\right) ,  \label{gamma1}
\end{equation}%
where the ($k$ dependent) $z_{i}$ and $\alpha _{i}$ are determined by 
requesting that Eq. (\ref{bloch}) is an eigenstate of $H_{k}^{0}$ with zero energy, which leads to 
a 2nd order polinomial in $z$, with roots $z_1,\;z_2$. $\zeta=sgn(\lambda \Delta_1)$ has been chosen to lead to $|z_i|<1$.
The normalization 
factor $N_{k},$ is determined from $\{\gamma _{k\uparrow },\gamma _{k\uparrow
}^{\dagger }\}=1$,
\begin{eqnarray}
N_{k}^{-2}&=&2\sum\limits_{j=1}^{N_{x}}\left\vert \sum\limits_{i=1}^{2}\alpha
_{i}z_{i}^{j-1}\right\vert ^{2}\nonumber \\
&\simeq& 2\left( \sum\limits_{i=1}^{2}\frac{%
\left\vert \alpha _{i}\right\vert ^{2}}{1-\left\vert z_{i}\right\vert ^{2}}+%
\frac{\alpha _{1}\bar{\alpha}_{2}}{1-z_{i}\bar{z}_{2}}+\frac{\bar{\alpha}%
_{1}\alpha _{2}}{1-\bar{z}_{1}z_{2}}\right) ,  \label{norm}
\end{eqnarray}%
where in the last equality, it has been assumed that $N_{x}$ is much larger
than the localization length of the zero mode.

Using time-reversal symmetry the corresponding solution for the Kramer's partner with spin down is obtained
\begin{equation}
\gamma _{Lk\downarrow }=N_{k}\sum\limits_{i=1}^{2}\bar{\alpha}_{i}
\sum\limits_{j=1}^{N_{x}}\bar{z}_{i}^{j-1}\left( c_{kj\downarrow}
+i\zeta c_{-kj\uparrow }^{\dagger }\right),  \label{gamma2}
\end{equation}
where we have used that $N_k$,  $\alpha_{i}$ and $z_i$ are even in $k$. Moving the parameters, this  continues to be valid
by continuity until $|z_{i}|=1$ is reached for one of the roots. At this
point the zero mode disappears [the normalization factor in Eq. (\ref{gamma1}%
) vanishes, see Eq. (\ref{norm})] and the system ceases to be topological.
Note that
\begin{equation}
\gamma _{Lk\sigma }^{\dagger }=-i\zeta \gamma _{L-k-\sigma }.  \label{rgam}
\end{equation}

So far, we have obtained exactly the zero-modes of the Hamiltonian disregarding 
$H_{k}^{\lambda }$. Using Eq. (\ref{gamma1}) we obtain for the complete
Hamiltonian
\begin{equation}
\lbrack \gamma _{Lk\uparrow },H]=\left( -2\lambda \sin k\right)
N_{k}\sum\limits_{i=1}^{2}\alpha
_{i}\sum\limits_{j=1}^{N_{x}}z_{i}^{j-1}\left( c_{kj\downarrow
}+i\zeta c_{-kj\uparrow }^{\dagger }\right) .  \label{comm}
\end{equation}%
The second member of this equation has a low-energy part proportional to $%
\gamma _{Lk\downarrow }$ and a high-energy part. In first order perturbation
theory in $H_{k}^{\lambda }$ we consider only the former part, which is
obtained anticommuting the second member with 
$\gamma_{Lk\downarrow }^{\dagger }$. The result is
\begin{equation}
\lbrack \gamma _{Lk\uparrow },H]=-2\lambda \rho _{k}e^{i\varphi _{k}}\sin
k\gamma _{Lk\downarrow }+...,  \label{comm2}
\end{equation}%
where ... denotes the high-energy part (a continuum of excited states) and $%
\rho _{k}$ and $\varphi _{k}$ are the modulus and phase of the complex number
\begin{equation}
\rho _{k}e^{i\varphi _{k}}=2N_{k}^{2}\sum\limits_{j=1}^{N_{x}}\left(
\sum\limits_{i=1}^{2}\alpha _{i}z_{i}^{j-1}\right) ^{2}\simeq
2N_{k}^{2}\left( \sum\limits_{i=1}^{2}\frac{\alpha _{i}^{2}}{1-z_{i}^{2}}+%
\frac{2\alpha _{1}\alpha _{2}}{1-z_{1}z_{2}}\right) ,  \label{rhodel}
\end{equation}%
where in the last equality, it has been assumed that $N_{x}$ is much larger
than the localization length of the zero mode. Note that all quantities in
Eq. (\ref{rhodel}) are even in $k$.

Using Eq. (\ref{comm2}) we obtain the eigenmodes
\begin{eqnarray}
\eta _{Lk\pm }&=&\frac{1}{\sqrt{2}}\left( \gamma _{Lk\uparrow }\pm e^{i\varphi
_{k}}\gamma _{Lk\downarrow }\right) ,\text{ }[\eta _{Lk\pm },H]=\pm \varepsilon_{\lambda,k},\nonumber \\%
\varepsilon_{\lambda,k}&=&-2\lambda \rho _{k}\sin k  \label{etaleft}
\end{eqnarray}%
Under time reversal $K$ these operators transform as
\begin{equation}
K\eta _{Lk\pm }K^{\dagger }=\mp e^{-i\varphi _{k}}\eta _{L-k\mp }.
\label{keta}
\end{equation}
Using Eqs. (\ref{rgam}) the following property is easily proved
\begin{equation}
\eta _{L-k\pm }^{\dagger }=\mp i\zeta e^{-i\varphi _{k}}\eta _{Lk\pm }.
\label{etac}
\end{equation}

The resulting energies are in excellent agreement with numerical results for
parameters well inside the topological region, for which $\rho _{k}\sim 1$,
for example $t=1$, $\mu =2$, $\Delta _{0}=4$, $\Delta _{1}=2.2$, $\lambda =7$
and $k$ near $\pi $. If the largest $\left\vert z_{i}\right\vert $
approaches 1, $\rho _{k}$ is small and the results lose accuracy . The
dependence of $\rho _{k}$ with $k$ is important. In general, the system is
topological for small $\xi _{k}$ and $\Delta _{k}$ [see Eqs. (\ref{xik})] 
\cite{zhang2013time}, a condition difficult to satisfy for all $k$ except for very
small $t$ and $\Delta _{0}$.

The low-energy states with important amplitude for sites near $j=N_{x}$ can
be obtained from those derived above by reflection ($j$ is interchanged with 
$N_{x}+1-j$) and complex conjugation (as done before \cite{aligia2018entangled} and
confirmed numerically). Then we have
\begin{eqnarray}
\gamma _{Rk\uparrow } &=&N_{k}\sum\limits_{i=1}^{2}\bar{\alpha}%
_{i}\sum\limits_{j=1}^{N_{x}}\bar{z}_{i}^{N_{x}-j}\left( c_{kj\uparrow
}-i\zeta c_{-kj\downarrow }^{\dagger }\right) ,  \nonumber \\
\gamma _{Rk\downarrow } &=&N_{k}\sum\limits_{i=1}^{2}\alpha
_{i}\sum\limits_{j=1}^{N_{x}}z_{i}^{N_{x}-j}\left( c_{kj\downarrow
}-i\zeta c_{-kj\uparrow }^{\dagger }\right) ,  \nonumber \\
\eta _{Rk\pm } &=&\frac{1}{\sqrt{2}}\left( \gamma _{Rk\uparrow }\pm
e^{-i\varphi _{k}}\gamma _{Rk\downarrow }\right) ,  \label{ger}
\end{eqnarray}%
with the following properties similar to Eqs. (\ref{rgam}) and (\ref{etac})
\begin{equation}
\gamma _{Rk\sigma }^{\dagger }=i\zeta \gamma _{R-k-\sigma }\text{, }\;\;\; \eta _{R-k\pm
}^{\dagger }=\pm i\zeta e^{i\varphi _{k}}\eta _{Rk\pm }.  \label{pro}
\end{equation}

\section{Derivation of the effective Hamiltonian for the TRITOPS-S junction} \label{appb}
We consider the low-energy effective Hamiltonian for the TRITOPS (S2) in Eq. (\ref{junction})  
\begin{equation}
H_{S2}=\sum\limits_{k>0} \varepsilon_{\lambda, k}\left( \eta _{k+}^{\dagger }\eta _{k+}-\eta
_{k-}^{\dagger }\eta _{k-}\right) ,
\label{hl}
\end{equation}%
with $\varepsilon_{\lambda,k}$ given by Eq. (\ref{ek}). The non-topological superconductor (S)  is modeled by Eq. (\ref{hsup}).
We substitute Eqs. (\ref{cend}) to (\ref{wnu}) in the operators of the TRITOPS side while for the S-side we use
\begin{equation}
c_{k\uparrow }^{\dagger }=\frac{1}{\sqrt{2}}\left( d_{k+}^{\dagger }+d
_{k-}^{\dagger }\right) ,\text{ }c_{k\downarrow }^{\dagger }=\frac{1}{\sqrt{2%
}}\left( d_{-k+}-d_{-k-}\right) .  \label{lefts}
\end{equation}
The effective  phase-dependent tunneling Hamiltonian for the junction, obtained after adding the contributions of $k$ and $-k$, reads
\begin{widetext}
\begin{eqnarray}
\frac{\sqrt{2}H_{\rm J}^{k}}{t_{J}} &=&\eta _{Lk+}^{\dagger }[w_{k}g_{-}(\phi )d_{k+}+w _{k}g_{+}(\phi )d_{k-}+
\bar{w}_{k}e^{i\varphi _{k}}g_{-}(\phi )d_{-k+}^{\dagger }-\bar{w}_{k}e^{i\varphi _{k}}g_{+}(\phi )d_{-k-}^{\dagger }]  \nonumber \\
&&+\eta _{Lk-}^{\dagger }[w_{k}g_{-}(\phi )d_{k+}+w_{k}g_{+}(\phi )d_{k-}-\bar{w}_{k}e^{i\varphi _{k}}g_{-}(\phi )d_{-k+}^{\dagger }+\bar{w}_{k}e^{i\varphi _{k}}g_{+}(\phi )d_{-k-}^{\dagger }]  +\text{H.c.,}  \label{hmix}
\end{eqnarray}%
\end{widetext}
where 
\begin{equation}
g_{\pm }(\phi )=e^{-i\phi /2}\pm i \zeta e^{i\phi /2}. \label{fphi}
\end{equation}
The fermionic degrees of freedom of S can be "integrated-out" by treating $H_{\rm J}^{k}$ in second-order of perturbation theory. The result is the effective Hamiltonian for the TRITOPS edge given in Eq. (\ref{zkm-s}).


\begin{thebibliography}{69}%
\makeatletter
\providecommand \@ifxundefined [1]{%
 \@ifx{#1\undefined}
}%
\providecommand \@ifnum [1]{%
 \ifnum #1\expandafter \@firstoftwo
 \else \expandafter \@secondoftwo
 \fi
}%
\providecommand \@ifx [1]{%
 \ifx #1\expandafter \@firstoftwo
 \else \expandafter \@secondoftwo
 \fi
}%
\providecommand \natexlab [1]{#1}%
\providecommand \enquote  [1]{``#1''}%
\providecommand \bibnamefont  [1]{#1}%
\providecommand \bibfnamefont [1]{#1}%
\providecommand \citenamefont [1]{#1}%
\providecommand \href@noop [0]{\@secondoftwo}%
\providecommand \href [0]{\begingroup \@sanitize@url \@href}%
\providecommand \@href[1]{\@@startlink{#1}\@@href}%
\providecommand \@@href[1]{\endgroup#1\@@endlink}%
\providecommand \@sanitize@url [0]{\catcode `\\12\catcode `\$12\catcode
  `\&12\catcode `\#12\catcode `\^12\catcode `\_12\catcode `\%12\relax}%
\providecommand \@@startlink[1]{}%
\providecommand \@@endlink[0]{}%
\providecommand \url  [0]{\begingroup\@sanitize@url \@url }%
\providecommand \@url [1]{\endgroup\@href {#1}{\urlprefix }}%
\providecommand \urlprefix  [0]{URL }%
\providecommand \Eprint [0]{\href }%
\providecommand \doibase [0]{https://doi.org/}%
\providecommand \selectlanguage [0]{\@gobble}%
\providecommand \bibinfo  [0]{\@secondoftwo}%
\providecommand \bibfield  [0]{\@secondoftwo}%
\providecommand \translation [1]{[#1]}%
\providecommand \BibitemOpen [0]{}%
\providecommand \bibitemStop [0]{}%
\providecommand \bibitemNoStop [0]{.\EOS\space}%
\providecommand \EOS [0]{\spacefactor3000\relax}%
\providecommand \BibitemShut  [1]{\csname bibitem#1\endcsname}%
\let\auto@bib@innerbib\@empty
\bibitem [{\citenamefont {Bernevig}(2013)}]{bernevig2013topological}%
  \BibitemOpen
  \bibfield  {author} {\bibinfo {author} {\bibfnamefont {B.~A.}\ \bibnamefont
  {Bernevig}},\ }\bibfield  {title} {\bibinfo {title} {Topological insulators
  and topological superconductors}\ }(\bibinfo  {publisher} {Princeton
  university press},\ \bibinfo {year} {2013})\BibitemShut {NoStop}%
\bibitem [{\citenamefont {Kitaev}(2001)}]{kitaev2001unpaired}%
  \BibitemOpen
  \bibfield  {author} {\bibinfo {author} {\bibfnamefont {A.~Y.}\ \bibnamefont
  {Kitaev}},\ }\bibfield  {title} {\bibinfo {title} {Unpaired majorana fermions
  in quantum wires},\ }\href@noop {} {\bibfield  {journal} {\bibinfo  {journal}
  {Physics-Uspekhi}\ }\textbf {\bibinfo {volume} {44}},\ \bibinfo {pages} {131}
  (\bibinfo {year} {2001})}\BibitemShut {NoStop}%
\bibitem [{\citenamefont {Freedman}\ \emph {et~al.}(2002)\citenamefont
  {Freedman}, \citenamefont {Larsen},\ and\ \citenamefont
  {Wang}}]{freedman2002modular}%
  \BibitemOpen
  \bibfield  {author} {\bibinfo {author} {\bibfnamefont {M.~H.}\ \bibnamefont
  {Freedman}}, \bibinfo {author} {\bibfnamefont {M.}~\bibnamefont {Larsen}},\
  and\ \bibinfo {author} {\bibfnamefont {Z.}~\bibnamefont {Wang}},\ }\bibfield
  {title} {\bibinfo {title} {A modular functor which is universal for quantum
  computation},\ }\href@noop {} {\bibfield  {journal} {\bibinfo  {journal}
  {Communications in Mathematical Physics}\ }\textbf {\bibinfo {volume}
  {227}},\ \bibinfo {pages} {605} (\bibinfo {year} {2002})}\BibitemShut
  {NoStop}%
\bibitem [{\citenamefont {Kitaev}(2003)}]{kitaev2003fault}%
  \BibitemOpen
  \bibfield  {author} {\bibinfo {author} {\bibfnamefont {A.~Y.}\ \bibnamefont
  {Kitaev}},\ }\bibfield  {title} {\bibinfo {title} {Fault-tolerant quantum
  computation by anyons},\ }\href@noop {} {\bibfield  {journal} {\bibinfo
  {journal} {Annals of Physics}\ }\textbf {\bibinfo {volume} {303}},\ \bibinfo
  {pages} {2} (\bibinfo {year} {2003})}\BibitemShut {NoStop}%
\bibitem [{\citenamefont {Lutchyn}\ \emph {et~al.}(2010)\citenamefont
  {Lutchyn}, \citenamefont {Sau},\ and\ \citenamefont {DasSarma}}]{wires1}%
  \BibitemOpen
  \bibfield  {author} {\bibinfo {author} {\bibfnamefont {R.~M.}\ \bibnamefont
  {Lutchyn}}, \bibinfo {author} {\bibfnamefont {J.~D.}\ \bibnamefont {Sau}},\
  and\ \bibinfo {author} {\bibfnamefont {S.}~\bibnamefont {DasSarma}},\
  }\bibfield  {title} {\bibinfo {title} {Majorana fermions and a topological
  phase transition in semiconductor-superconductor heterostructures},\
  }\href@noop {} {\bibfield  {journal} {\bibinfo  {journal} {Phys. Rev. Lett.}\
  }\textbf {\bibinfo {volume} {105}},\ \bibinfo {pages} {077001} (\bibinfo
  {year} {2010})}\BibitemShut {NoStop}%
\bibitem [{\citenamefont {Oreg}\ \emph {et~al.}(2010)\citenamefont {Oreg},
  \citenamefont {Refael},\ and\ \citenamefont {von Oppen}}]{wires2}%
  \BibitemOpen
  \bibfield  {author} {\bibinfo {author} {\bibfnamefont {Y.}~\bibnamefont
  {Oreg}}, \bibinfo {author} {\bibfnamefont {G.}~\bibnamefont {Refael}},\ and\
  \bibinfo {author} {\bibfnamefont {F.}~\bibnamefont {von Oppen}},\ }\bibfield
  {title} {\bibinfo {title} {Helical liquids and majorana bound states in
  quantum wires},\ }\href@noop {} {\bibfield  {journal} {\bibinfo  {journal}
  {Phys. Rev. Lett.}\ }\textbf {\bibinfo {volume} {105}},\ \bibinfo {pages}
  {177002} (\bibinfo {year} {2010})}\BibitemShut {NoStop}%
\bibitem [{\citenamefont {Mourik}\ \emph {et~al.}(2012)\citenamefont {Mourik},
  \citenamefont {Zuo}, \citenamefont {Frolov}, \citenamefont {Plissard},
  \citenamefont {Bakkers},\ and\ \citenamefont
  {Kouwenhoven}}]{mourik2012signatures}%
  \BibitemOpen
  \bibfield  {author} {\bibinfo {author} {\bibfnamefont {V.}~\bibnamefont
  {Mourik}}, \bibinfo {author} {\bibfnamefont {K.}~\bibnamefont {Zuo}},
  \bibinfo {author} {\bibfnamefont {S.~M.}\ \bibnamefont {Frolov}}, \bibinfo
  {author} {\bibfnamefont {S.}~\bibnamefont {Plissard}}, \bibinfo {author}
  {\bibfnamefont {E.~P.}\ \bibnamefont {Bakkers}},\ and\ \bibinfo {author}
  {\bibfnamefont {L.~P.}\ \bibnamefont {Kouwenhoven}},\ }\bibfield  {title}
  {\bibinfo {title} {Signatures of majorana fermions in hybrid
  superconductor-semiconductor nanowire devices},\ }\href@noop {} {\bibfield
  {journal} {\bibinfo  {journal} {Science}\ }\textbf {\bibinfo {volume}
  {336}},\ \bibinfo {pages} {1003} (\bibinfo {year} {2012})}\BibitemShut
  {NoStop}%
\bibitem [{\citenamefont {Rokhinson}\ \emph {et~al.}(2012)\citenamefont
  {Rokhinson}, \citenamefont {Liu},\ and\ \citenamefont
  {Furdyna}}]{rokhinson2012fractional}%
  \BibitemOpen
  \bibfield  {author} {\bibinfo {author} {\bibfnamefont {L.~P.}\ \bibnamefont
  {Rokhinson}}, \bibinfo {author} {\bibfnamefont {X.}~\bibnamefont {Liu}},\
  and\ \bibinfo {author} {\bibfnamefont {J.~K.}\ \bibnamefont {Furdyna}},\
  }\bibfield  {title} {\bibinfo {title} {The fractional ac josephson effect in
  a semiconductor--superconductor nanowire as a signature of majorana
  particles},\ }\href@noop {} {\bibfield  {journal} {\bibinfo  {journal}
  {Nature Physics}\ }\textbf {\bibinfo {volume} {8}},\ \bibinfo {pages} {795}
  (\bibinfo {year} {2012})}\BibitemShut {NoStop}%
\bibitem [{\citenamefont {Das}\ \emph {et~al.}(2012)\citenamefont {Das},
  \citenamefont {Ronen}, \citenamefont {Most}, \citenamefont {Oreg},
  \citenamefont {Heiblum},\ and\ \citenamefont {Shtrikman}}]{das2012zero}%
  \BibitemOpen
  \bibfield  {author} {\bibinfo {author} {\bibfnamefont {A.}~\bibnamefont
  {Das}}, \bibinfo {author} {\bibfnamefont {Y.}~\bibnamefont {Ronen}}, \bibinfo
  {author} {\bibfnamefont {Y.}~\bibnamefont {Most}}, \bibinfo {author}
  {\bibfnamefont {Y.}~\bibnamefont {Oreg}}, \bibinfo {author} {\bibfnamefont
  {M.}~\bibnamefont {Heiblum}},\ and\ \bibinfo {author} {\bibfnamefont
  {H.}~\bibnamefont {Shtrikman}},\ }\bibfield  {title} {\bibinfo {title}
  {Zero-bias peaks and splitting in an al--inas nanowire topological
  superconductor as a signature of majorana fermions},\ }\href@noop {}
  {\bibfield  {journal} {\bibinfo  {journal} {Nature Physics}\ }\textbf
  {\bibinfo {volume} {8}},\ \bibinfo {pages} {887} (\bibinfo {year}
  {2012})}\BibitemShut {NoStop}%
\bibitem [{\citenamefont {Albrecht}\ \emph {et~al.}(2016)\citenamefont
  {Albrecht}, \citenamefont {Higginbotham}, \citenamefont {Madsen},
  \citenamefont {Kuemmeth}, \citenamefont {Jespersen}, \citenamefont
  {Nyg{\aa}rd}, \citenamefont {Krogstrup},\ and\ \citenamefont
  {Marcus}}]{albrecht2016exponential}%
  \BibitemOpen
  \bibfield  {author} {\bibinfo {author} {\bibfnamefont {S.~M.}\ \bibnamefont
  {Albrecht}}, \bibinfo {author} {\bibfnamefont {A.~P.}\ \bibnamefont
  {Higginbotham}}, \bibinfo {author} {\bibfnamefont {M.}~\bibnamefont
  {Madsen}}, \bibinfo {author} {\bibfnamefont {F.}~\bibnamefont {Kuemmeth}},
  \bibinfo {author} {\bibfnamefont {T.~S.}\ \bibnamefont {Jespersen}}, \bibinfo
  {author} {\bibfnamefont {J.}~\bibnamefont {Nyg{\aa}rd}}, \bibinfo {author}
  {\bibfnamefont {P.}~\bibnamefont {Krogstrup}},\ and\ \bibinfo {author}
  {\bibfnamefont {C.}~\bibnamefont {Marcus}},\ }\bibfield  {title} {\bibinfo
  {title} {Exponential protection of zero modes in majorana islands},\
  }\href@noop {} {\bibfield  {journal} {\bibinfo  {journal} {Nature}\ }\textbf
  {\bibinfo {volume} {531}},\ \bibinfo {pages} {206} (\bibinfo {year}
  {2016})}\BibitemShut {NoStop}%
\bibitem [{\citenamefont {Deng}\ \emph {et~al.}(2012)\citenamefont {Deng},
  \citenamefont {Viola},\ and\ \citenamefont {Ortiz}}]{deng2012majorana}%
  \BibitemOpen
  \bibfield  {author} {\bibinfo {author} {\bibfnamefont {S.}~\bibnamefont
  {Deng}}, \bibinfo {author} {\bibfnamefont {L.}~\bibnamefont {Viola}},\ and\
  \bibinfo {author} {\bibfnamefont {G.}~\bibnamefont {Ortiz}},\ }\bibfield
  {title} {\bibinfo {title} {Majorana modes in time-reversal invariant s-wave
  topological superconductors},\ }\href@noop {} {\bibfield  {journal} {\bibinfo
   {journal} {Phys. Rev. Lett.}\ }\textbf {\bibinfo {volume} {108}},\ \bibinfo
  {pages} {036803} (\bibinfo {year} {2012})}\BibitemShut {NoStop}%
\bibitem [{\citenamefont {Nadj-Perge}\ \emph {et~al.}(2014)\citenamefont
  {Nadj-Perge}, \citenamefont {Drozdov}, \citenamefont {Li}, \citenamefont
  {Chen}, \citenamefont {Jeon}, \citenamefont {Seo}, \citenamefont {MacDonald},
  \citenamefont {Bernevig},\ and\ \citenamefont {Yazdani}}]{yazdani}%
  \BibitemOpen
  \bibfield  {author} {\bibinfo {author} {\bibfnamefont {S.}~\bibnamefont
  {Nadj-Perge}}, \bibinfo {author} {\bibfnamefont {I.~K.}\ \bibnamefont
  {Drozdov}}, \bibinfo {author} {\bibfnamefont {J.}~\bibnamefont {Li}},
  \bibinfo {author} {\bibfnamefont {H.}~\bibnamefont {Chen}}, \bibinfo {author}
  {\bibfnamefont {S.}~\bibnamefont {Jeon}}, \bibinfo {author} {\bibfnamefont
  {J.}~\bibnamefont {Seo}}, \bibinfo {author} {\bibfnamefont {A.~H.}\
  \bibnamefont {MacDonald}}, \bibinfo {author} {\bibfnamefont {B.~A.}\
  \bibnamefont {Bernevig}},\ and\ \bibinfo {author} {\bibfnamefont
  {A.}~\bibnamefont {Yazdani}},\ }\bibfield  {title} {\bibinfo {title}
  {Observation of majorana fermions in ferromagnetic atomic chains on a
  superconductor},\ }\href@noop {} {\bibfield  {journal} {\bibinfo  {journal}
  {Science}\ }\textbf {\bibinfo {volume} {346}},\ \bibinfo {pages} {602}
  (\bibinfo {year} {2014})}\BibitemShut {NoStop}%
\bibitem [{\citenamefont {Kim}\ \emph {et~al.}(2018)\citenamefont {Kim},
  \citenamefont {Palacio-Morales}, \citenamefont {Posske}, \citenamefont
  {R{\'o}zsa}, \citenamefont {Palot{\'a}s}, \citenamefont {Szunyogh},
  \citenamefont {Thorwart},\ and\ \citenamefont {Wiesendanger}}]{wiesendanger}%
  \BibitemOpen
  \bibfield  {author} {\bibinfo {author} {\bibfnamefont {H.}~\bibnamefont
  {Kim}}, \bibinfo {author} {\bibfnamefont {A.}~\bibnamefont
  {Palacio-Morales}}, \bibinfo {author} {\bibfnamefont {T.}~\bibnamefont
  {Posske}}, \bibinfo {author} {\bibfnamefont {L.}~\bibnamefont {R{\'o}zsa}},
  \bibinfo {author} {\bibfnamefont {K.}~\bibnamefont {Palot{\'a}s}}, \bibinfo
  {author} {\bibfnamefont {L.}~\bibnamefont {Szunyogh}}, \bibinfo {author}
  {\bibfnamefont {M.}~\bibnamefont {Thorwart}},\ and\ \bibinfo {author}
  {\bibfnamefont {R.}~\bibnamefont {Wiesendanger}},\ }\bibfield  {title}
  {\bibinfo {title} {Toward tailoring majorana bound states in artificially
  constructed magnetic atom chains on elemental superconductors},\ }\href@noop
  {} {\bibfield  {journal} {\bibinfo  {journal} {Science Advances}\ }\textbf
  {\bibinfo {volume} {4}},\ \bibinfo {pages} {eaar5251} (\bibinfo {year}
  {2018})}\BibitemShut {NoStop}%
\bibitem [{\citenamefont {Ruby}\ \emph {et~al.}(2015)\citenamefont {Ruby},
  \citenamefont {Pientka}, \citenamefont {Peng}, \citenamefont {von Oppen},
  \citenamefont {Heinrich},\ and\ \citenamefont {Franke}}]{ruby}%
  \BibitemOpen
  \bibfield  {author} {\bibinfo {author} {\bibfnamefont {M.}~\bibnamefont
  {Ruby}}, \bibinfo {author} {\bibfnamefont {F.}~\bibnamefont {Pientka}},
  \bibinfo {author} {\bibfnamefont {Y.}~\bibnamefont {Peng}}, \bibinfo {author}
  {\bibfnamefont {F.}~\bibnamefont {von Oppen}}, \bibinfo {author}
  {\bibfnamefont {B.~W.}\ \bibnamefont {Heinrich}},\ and\ \bibinfo {author}
  {\bibfnamefont {K.~J.}\ \bibnamefont {Franke}},\ }\bibfield  {title}
  {\bibinfo {title} {End states and subgap structure in proximity-coupled
  chains of magnetic adatoms},\ }\href@noop {} {\bibfield  {journal} {\bibinfo
  {journal} {Phys. Rev. Lett.}\ }\textbf {\bibinfo {volume} {115}},\ \bibinfo
  {pages} {197204} (\bibinfo {year} {2015})}\BibitemShut {NoStop}%
\bibitem [{\citenamefont {Zhang}\ \emph {et~al.}(2018)\citenamefont {Zhang},
  \citenamefont {Yaji}, \citenamefont {Hashimoto}, \citenamefont {Ota},
  \citenamefont {Kondo}, \citenamefont {Okazaki}, \citenamefont {Wang},
  \citenamefont {Wen}, \citenamefont {Gu}, \citenamefont {Ding} \emph
  {et~al.}}]{iron}%
  \BibitemOpen
  \bibfield  {author} {\bibinfo {author} {\bibfnamefont {P.}~\bibnamefont
  {Zhang}}, \bibinfo {author} {\bibfnamefont {K.}~\bibnamefont {Yaji}},
  \bibinfo {author} {\bibfnamefont {T.}~\bibnamefont {Hashimoto}}, \bibinfo
  {author} {\bibfnamefont {Y.}~\bibnamefont {Ota}}, \bibinfo {author}
  {\bibfnamefont {T.}~\bibnamefont {Kondo}}, \bibinfo {author} {\bibfnamefont
  {K.}~\bibnamefont {Okazaki}}, \bibinfo {author} {\bibfnamefont
  {Z.}~\bibnamefont {Wang}}, \bibinfo {author} {\bibfnamefont {J.}~\bibnamefont
  {Wen}}, \bibinfo {author} {\bibfnamefont {G.}~\bibnamefont {Gu}}, \bibinfo
  {author} {\bibfnamefont {H.}~\bibnamefont {Ding}}, \emph {et~al.},\
  }\bibfield  {title} {\bibinfo {title} {Observation of topological
  superconductivity on the surface of an iron-based superconductor},\
  }\href@noop {} {\bibfield  {journal} {\bibinfo  {journal} {Science}\ }\textbf
  {\bibinfo {volume} {360}},\ \bibinfo {pages} {182} (\bibinfo {year}
  {2018})}\BibitemShut {NoStop}%
\bibitem [{\citenamefont {Fu}\ and\ \citenamefont
  {Kane}(2008)}]{fu2008superconducting}%
  \BibitemOpen
  \bibfield  {author} {\bibinfo {author} {\bibfnamefont {L.}~\bibnamefont
  {Fu}}\ and\ \bibinfo {author} {\bibfnamefont {C.~L.}\ \bibnamefont {Kane}},\
  }\bibfield  {title} {\bibinfo {title} {Superconducting proximity effect and
  majorana fermions at the surface of a topological insulator},\ }\href@noop {}
  {\bibfield  {journal} {\bibinfo  {journal} {Phys. Rev. Lett.}\ }\textbf
  {\bibinfo {volume} {100}},\ \bibinfo {pages} {096407} (\bibinfo {year}
  {2008})}\BibitemShut {NoStop}%
\bibitem [{\citenamefont {Fu}\ and\ \citenamefont
  {Kane}(2009)}]{fu2009josephson}%
  \BibitemOpen
  \bibfield  {author} {\bibinfo {author} {\bibfnamefont {L.}~\bibnamefont
  {Fu}}\ and\ \bibinfo {author} {\bibfnamefont {C.~L.}\ \bibnamefont {Kane}},\
  }\bibfield  {title} {\bibinfo {title} {Josephson current and noise at a
  superconductor/quantum-spin-hall-insulator/superconductor junction},\
  }\href@noop {} {\bibfield  {journal} {\bibinfo  {journal} {Physical Review
  B}\ }\textbf {\bibinfo {volume} {79}},\ \bibinfo {pages} {161408(R)}
  (\bibinfo {year} {2009})}\BibitemShut {NoStop}%
\bibitem [{\citenamefont {Qi}\ and\ \citenamefont {Zhang}(2011)}]{zhang}%
  \BibitemOpen
  \bibfield  {author} {\bibinfo {author} {\bibfnamefont {X.-L.}\ \bibnamefont
  {Qi}}\ and\ \bibinfo {author} {\bibfnamefont {S.-C.}\ \bibnamefont {Zhang}},\
  }\bibfield  {title} {\bibinfo {title} {Topological insulators and
  superconductors},\ }\href@noop {} {\bibfield  {journal} {\bibinfo  {journal}
  {Reviews of Modern Physics}\ }\textbf {\bibinfo {volume} {83}},\ \bibinfo
  {pages} {1057} (\bibinfo {year} {2011})}\BibitemShut {NoStop}%
\bibitem [{\citenamefont {Alicea}(2012)}]{alicea2012new}%
  \BibitemOpen
  \bibfield  {author} {\bibinfo {author} {\bibfnamefont {J.}~\bibnamefont
  {Alicea}},\ }\bibfield  {title} {\bibinfo {title} {New directions in the
  pursuit of majorana fermions in solid state systems},\ }\href@noop {}
  {\bibfield  {journal} {\bibinfo  {journal} {Reports on progress in physics}\
  }\textbf {\bibinfo {volume} {75}},\ \bibinfo {pages} {076501} (\bibinfo
  {year} {2012})}\BibitemShut {NoStop}%
\bibitem [{\citenamefont {Aguado}(2017)}]{aguado2017majorana}%
  \BibitemOpen
  \bibfield  {author} {\bibinfo {author} {\bibfnamefont {R.}~\bibnamefont
  {Aguado}},\ }\bibfield  {title} {\bibinfo {title} {Majorana quasiparticles in
  condensed matter},\ }\href@noop {} {\bibfield  {journal} {\bibinfo  {journal}
  {La Rivista del Nuovo Cimento}\ }\textbf {\bibinfo {volume} {40}},\ \bibinfo
  {pages} {523} (\bibinfo {year} {2017})}\BibitemShut {NoStop}%
\bibitem [{\citenamefont {Flensberg}\ \emph {et~al.}(2021)\citenamefont
  {Flensberg}, \citenamefont {von Oppen},\ and\ \citenamefont
  {Stern}}]{flensberg2021engineered}%
  \BibitemOpen
  \bibfield  {author} {\bibinfo {author} {\bibfnamefont {K.}~\bibnamefont
  {Flensberg}}, \bibinfo {author} {\bibfnamefont {F.}~\bibnamefont {von
  Oppen}},\ and\ \bibinfo {author} {\bibfnamefont {A.}~\bibnamefont {Stern}},\
  }\bibfield  {title} {\bibinfo {title} {Engineered platforms for topological
  superconductivity and majorana zero modes},\ }\href@noop {} {\bibfield
  {journal} {\bibinfo  {journal} {Nature Reviews Materials}\ }\textbf {\bibinfo
  {volume} {6}},\ \bibinfo {pages} {944} (\bibinfo {year} {2021})}\BibitemShut
  {NoStop}%
\bibitem [{\citenamefont {Ryu}\ \emph {et~al.}(2010)\citenamefont {Ryu},
  \citenamefont {Schnyder}, \citenamefont {Furusaki},\ and\ \citenamefont
  {Ludwig}}]{ryu2010topological}%
  \BibitemOpen
  \bibfield  {author} {\bibinfo {author} {\bibfnamefont {S.}~\bibnamefont
  {Ryu}}, \bibinfo {author} {\bibfnamefont {A.~P.}\ \bibnamefont {Schnyder}},
  \bibinfo {author} {\bibfnamefont {A.}~\bibnamefont {Furusaki}},\ and\
  \bibinfo {author} {\bibfnamefont {A.~W.}\ \bibnamefont {Ludwig}},\ }\bibfield
   {title} {\bibinfo {title} {Topological insulators and superconductors:
  tenfold way and dimensional hierarchy},\ }\href@noop {} {\bibfield  {journal}
  {\bibinfo  {journal} {New Journal of Physics}\ }\textbf {\bibinfo {volume}
  {12}},\ \bibinfo {pages} {065010} (\bibinfo {year} {2010})}\BibitemShut
  {NoStop}%
\bibitem [{\citenamefont {Qi}\ \emph {et~al.}(2009)\citenamefont {Qi},
  \citenamefont {Hughes}, \citenamefont {Raghu},\ and\ \citenamefont
  {Zhang}}]{qi2009time}%
  \BibitemOpen
  \bibfield  {author} {\bibinfo {author} {\bibfnamefont {X.-L.}\ \bibnamefont
  {Qi}}, \bibinfo {author} {\bibfnamefont {T.~L.}\ \bibnamefont {Hughes}},
  \bibinfo {author} {\bibfnamefont {S.}~\bibnamefont {Raghu}},\ and\ \bibinfo
  {author} {\bibfnamefont {S.-C.}\ \bibnamefont {Zhang}},\ }\bibfield  {title}
  {\bibinfo {title} {Time-reversal-invariant topological superconductors and
  superfluids in two and three dimensions},\ }\href@noop {} {\bibfield
  {journal} {\bibinfo  {journal} {Phys. Rev. Lett.}\ }\textbf {\bibinfo
  {volume} {102}},\ \bibinfo {pages} {187001} (\bibinfo {year}
  {2009})}\BibitemShut {NoStop}%
\bibitem [{\citenamefont {Dumitrescu}\ and\ \citenamefont
  {Tewari}(2013)}]{dumitrescu2013topological}%
  \BibitemOpen
  \bibfield  {author} {\bibinfo {author} {\bibfnamefont {E.}~\bibnamefont
  {Dumitrescu}}\ and\ \bibinfo {author} {\bibfnamefont {S.}~\bibnamefont
  {Tewari}},\ }\bibfield  {title} {\bibinfo {title} {Topological properties of
  the time-reversal-symmetric kitaev chain and applications to organic
  superconductors},\ }\href@noop {} {\bibfield  {journal} {\bibinfo  {journal}
  {Physical Review B}\ }\textbf {\bibinfo {volume} {88}},\ \bibinfo {pages}
  {220505(R)} (\bibinfo {year} {2013})}\BibitemShut {NoStop}%
\bibitem [{\citenamefont {Haim}\ \emph {et~al.}(2014)\citenamefont {Haim},
  \citenamefont {Keselman}, \citenamefont {Berg},\ and\ \citenamefont
  {Oreg}}]{haim2014time}%
  \BibitemOpen
  \bibfield  {author} {\bibinfo {author} {\bibfnamefont {A.}~\bibnamefont
  {Haim}}, \bibinfo {author} {\bibfnamefont {A.}~\bibnamefont {Keselman}},
  \bibinfo {author} {\bibfnamefont {E.}~\bibnamefont {Berg}},\ and\ \bibinfo
  {author} {\bibfnamefont {Y.}~\bibnamefont {Oreg}},\ }\bibfield  {title}
  {\bibinfo {title} {Time-reversal-invariant topological superconductivity
  induced by repulsive interactions in quantum wires},\ }\href@noop {}
  {\bibfield  {journal} {\bibinfo  {journal} {Physical Review B}\ }\textbf
  {\bibinfo {volume} {89}},\ \bibinfo {pages} {220504(R)} (\bibinfo {year}
  {2014})}\BibitemShut {NoStop}%
\bibitem [{\citenamefont {Tanaka}\ \emph {et~al.}(2010)\citenamefont {Tanaka},
  \citenamefont {Mizuno}, \citenamefont {Yokoyama}, \citenamefont {Yada},\ and\
  \citenamefont {Sato}}]{tanaka-tritops}%
  \BibitemOpen
  \bibfield  {author} {\bibinfo {author} {\bibfnamefont {Y.}~\bibnamefont
  {Tanaka}}, \bibinfo {author} {\bibfnamefont {Y.}~\bibnamefont {Mizuno}},
  \bibinfo {author} {\bibfnamefont {T.}~\bibnamefont {Yokoyama}}, \bibinfo
  {author} {\bibfnamefont {K.}~\bibnamefont {Yada}},\ and\ \bibinfo {author}
  {\bibfnamefont {M.}~\bibnamefont {Sato}},\ }\bibfield  {title} {\bibinfo
  {title} {Anomalous andreev bound state in noncentrosymmetric
  superconductors},\ }\href {https://doi.org/10.1103/PhysRevLett.105.097002}
  {\bibfield  {journal} {\bibinfo  {journal} {Phys. Rev. Lett.}\ }\textbf
  {\bibinfo {volume} {105}},\ \bibinfo {pages} {097002} (\bibinfo {year}
  {2010})}\BibitemShut {NoStop}%
\bibitem [{\citenamefont {H.~Kwon}\ and\ \citenamefont
  {Yakovenko}(2004)}]{kwon}%
  \BibitemOpen
  \bibfield  {author} {\bibinfo {author} {\bibfnamefont {K.~S.}\ \bibnamefont
  {H.~Kwon}}\ and\ \bibinfo {author} {\bibfnamefont {V.}~\bibnamefont
  {Yakovenko}},\ }\bibfield  {title} {\bibinfo {title} {Fractional ac josephson
  effect in p-and d-wave superconductors},\ }\href@noop {} {\bibfield
  {journal} {\bibinfo  {journal} {Eur. Phys. J. B}\ }\textbf {\bibinfo {volume}
  {37}},\ \bibinfo {pages} {349} (\bibinfo {year} {2004})}\BibitemShut
  {NoStop}%
\bibitem [{\citenamefont {Fu}\ and\ \citenamefont {Berg}(2010)}]{fu2010odd}%
  \BibitemOpen
  \bibfield  {author} {\bibinfo {author} {\bibfnamefont {L.}~\bibnamefont
  {Fu}}\ and\ \bibinfo {author} {\bibfnamefont {E.}~\bibnamefont {Berg}},\
  }\bibfield  {title} {\bibinfo {title} {Odd-parity topological
  superconductors: theory and application to {Cu$_{x}$Bi$_{2}$Se$_{3}$}},\
  }\href@noop {} {\bibfield  {journal} {\bibinfo  {journal} {Phys. Rev. Lett.}\
  }\textbf {\bibinfo {volume} {105}},\ \bibinfo {pages} {097001} (\bibinfo
  {year} {2010})}\BibitemShut {NoStop}%
\bibitem [{\citenamefont {Scheurer}\ and\ \citenamefont
  {Schmalian}(2015)}]{scheurer2015topological}%
  \BibitemOpen
  \bibfield  {author} {\bibinfo {author} {\bibfnamefont {M.~S.}\ \bibnamefont
  {Scheurer}}\ and\ \bibinfo {author} {\bibfnamefont {J.}~\bibnamefont
  {Schmalian}},\ }\bibfield  {title} {\bibinfo {title} {Topological
  superconductivity and unconventional pairing in oxide interfaces},\
  }\href@noop {} {\bibfield  {journal} {\bibinfo  {journal} {Nature
  communications}\ }\textbf {\bibinfo {volume} {6}},\ \bibinfo {pages} {1}
  (\bibinfo {year} {2015})}\BibitemShut {NoStop}%
\bibitem [{\citenamefont {Wong}\ and\ \citenamefont
  {Law}(2012)}]{wong2012majorana}%
  \BibitemOpen
  \bibfield  {author} {\bibinfo {author} {\bibfnamefont {C.~L.~M.}\
  \bibnamefont {Wong}}\ and\ \bibinfo {author} {\bibfnamefont {K.~T.}\
  \bibnamefont {Law}},\ }\bibfield  {title} {\bibinfo {title} {Majorana kramers
  doublets in d x 2- y 2-wave superconductors with rashba spin-orbit
  coupling},\ }\href@noop {} {\bibfield  {journal} {\bibinfo  {journal}
  {Physical Review B}\ }\textbf {\bibinfo {volume} {86}},\ \bibinfo {pages}
  {184516} (\bibinfo {year} {2012})}\BibitemShut {NoStop}%
\bibitem [{\citenamefont {Zhang}\ \emph {et~al.}(2013)\citenamefont {Zhang},
  \citenamefont {Kane},\ and\ \citenamefont {Mele}}]{zhang2013time}%
  \BibitemOpen
  \bibfield  {author} {\bibinfo {author} {\bibfnamefont {F.}~\bibnamefont
  {Zhang}}, \bibinfo {author} {\bibfnamefont {C.~L.}\ \bibnamefont {Kane}},\
  and\ \bibinfo {author} {\bibfnamefont {E.~J.}\ \bibnamefont {Mele}},\
  }\bibfield  {title} {\bibinfo {title} {Time-reversal-invariant topological
  superconductivity and majorana kramers pairs},\ }\href@noop {} {\bibfield
  {journal} {\bibinfo  {journal} {Phys. Rev. Lett.}\ }\textbf {\bibinfo
  {volume} {111}},\ \bibinfo {pages} {056402} (\bibinfo {year}
  {2013})}\BibitemShut {NoStop}%
\bibitem [{\citenamefont {Keselman}\ \emph {et~al.}(2013)\citenamefont
  {Keselman}, \citenamefont {Fu}, \citenamefont {Stern},\ and\ \citenamefont
  {Berg}}]{keselman2013inducing}%
  \BibitemOpen
  \bibfield  {author} {\bibinfo {author} {\bibfnamefont {A.}~\bibnamefont
  {Keselman}}, \bibinfo {author} {\bibfnamefont {L.}~\bibnamefont {Fu}},
  \bibinfo {author} {\bibfnamefont {A.}~\bibnamefont {Stern}},\ and\ \bibinfo
  {author} {\bibfnamefont {E.}~\bibnamefont {Berg}},\ }\bibfield  {title}
  {\bibinfo {title} {Inducing time-reversal-invariant topological
  superconductivity and fermion parity pumping in quantum wires},\ }\href@noop
  {} {\bibfield  {journal} {\bibinfo  {journal} {Phys. Rev. Lett.}\ }\textbf
  {\bibinfo {volume} {111}},\ \bibinfo {pages} {116402} (\bibinfo {year}
  {2013})}\BibitemShut {NoStop}%
\bibitem [{\citenamefont {Haim}\ \emph {et~al.}(2016)\citenamefont {Haim},
  \citenamefont {W{\"o}lms}, \citenamefont {Berg}, \citenamefont {Oreg},\ and\
  \citenamefont {Flensberg}}]{haim2016interaction}%
  \BibitemOpen
  \bibfield  {author} {\bibinfo {author} {\bibfnamefont {A.}~\bibnamefont
  {Haim}}, \bibinfo {author} {\bibfnamefont {K.}~\bibnamefont {W{\"o}lms}},
  \bibinfo {author} {\bibfnamefont {E.}~\bibnamefont {Berg}}, \bibinfo {author}
  {\bibfnamefont {Y.}~\bibnamefont {Oreg}},\ and\ \bibinfo {author}
  {\bibfnamefont {K.}~\bibnamefont {Flensberg}},\ }\bibfield  {title} {\bibinfo
  {title} {Interaction-driven topological superconductivity in one dimension},\
  }\href@noop {} {\bibfield  {journal} {\bibinfo  {journal} {Physical Review
  B}\ }\textbf {\bibinfo {volume} {94}},\ \bibinfo {pages} {115124} (\bibinfo
  {year} {2016})}\BibitemShut {NoStop}%
\bibitem [{\citenamefont {Reeg}\ \emph {et~al.}(2017)\citenamefont {Reeg},
  \citenamefont {Schrade}, \citenamefont {Klinovaja},\ and\ \citenamefont
  {Loss}}]{reeg2017diii}%
  \BibitemOpen
  \bibfield  {author} {\bibinfo {author} {\bibfnamefont {C.}~\bibnamefont
  {Reeg}}, \bibinfo {author} {\bibfnamefont {C.}~\bibnamefont {Schrade}},
  \bibinfo {author} {\bibfnamefont {J.}~\bibnamefont {Klinovaja}},\ and\
  \bibinfo {author} {\bibfnamefont {D.}~\bibnamefont {Loss}},\ }\bibfield
  {title} {\bibinfo {title} {Diii topological superconductivity with emergent
  time-reversal symmetry},\ }\href@noop {} {\bibfield  {journal} {\bibinfo
  {journal} {Physical Review B}\ }\textbf {\bibinfo {volume} {96}},\ \bibinfo
  {pages} {161407(R)} (\bibinfo {year} {2017})}\BibitemShut {NoStop}%
\bibitem [{\citenamefont {Santos}\ \emph {et~al.}(2010)\citenamefont {Santos},
  \citenamefont {Neupert}, \citenamefont {Chamon},\ and\ \citenamefont
  {Mudry}}]{santos2010superconductivity}%
  \BibitemOpen
  \bibfield  {author} {\bibinfo {author} {\bibfnamefont {L.}~\bibnamefont
  {Santos}}, \bibinfo {author} {\bibfnamefont {T.}~\bibnamefont {Neupert}},
  \bibinfo {author} {\bibfnamefont {C.}~\bibnamefont {Chamon}},\ and\ \bibinfo
  {author} {\bibfnamefont {C.}~\bibnamefont {Mudry}},\ }\bibfield  {title}
  {\bibinfo {title} {Superconductivity on the surface of topological insulators
  and in two-dimensional noncentrosymmetric materials},\ }\href@noop {}
  {\bibfield  {journal} {\bibinfo  {journal} {Physical Review B}\ }\textbf
  {\bibinfo {volume} {81}},\ \bibinfo {pages} {184502} (\bibinfo {year}
  {2010})}\BibitemShut {NoStop}%
\bibitem [{\citenamefont {Klinovaja}\ \emph {et~al.}(2014)\citenamefont
  {Klinovaja}, \citenamefont {Yacoby},\ and\ \citenamefont
  {Loss}}]{klinovaja2014kramers}%
  \BibitemOpen
  \bibfield  {author} {\bibinfo {author} {\bibfnamefont {J.}~\bibnamefont
  {Klinovaja}}, \bibinfo {author} {\bibfnamefont {A.}~\bibnamefont {Yacoby}},\
  and\ \bibinfo {author} {\bibfnamefont {D.}~\bibnamefont {Loss}},\ }\bibfield
  {title} {\bibinfo {title} {Kramers pairs of majorana fermions and
  parafermions in fractional topological insulators},\ }\href@noop {}
  {\bibfield  {journal} {\bibinfo  {journal} {Physical Review B}\ }\textbf
  {\bibinfo {volume} {90}},\ \bibinfo {pages} {155447} (\bibinfo {year}
  {2014})}\BibitemShut {NoStop}%
\bibitem [{\citenamefont {Mellars}\ and\ \citenamefont
  {B{\'e}ri}(2016)}]{mellars2016signatures}%
  \BibitemOpen
  \bibfield  {author} {\bibinfo {author} {\bibfnamefont {E.}~\bibnamefont
  {Mellars}}\ and\ \bibinfo {author} {\bibfnamefont {B.}~\bibnamefont
  {B{\'e}ri}},\ }\bibfield  {title} {\bibinfo {title} {Signatures of
  time-reversal-invariant topological superconductivity in the josephson
  effect},\ }\href@noop {} {\bibfield  {journal} {\bibinfo  {journal} {Physical
  Review B}\ }\textbf {\bibinfo {volume} {94}},\ \bibinfo {pages} {174508}
  (\bibinfo {year} {2016})}\BibitemShut {NoStop}%
\bibitem [{\citenamefont {Parhizgar}\ and\ \citenamefont
  {Black-Schaffer}(2017)}]{parhizgar2017highly}%
  \BibitemOpen
  \bibfield  {author} {\bibinfo {author} {\bibfnamefont {F.}~\bibnamefont
  {Parhizgar}}\ and\ \bibinfo {author} {\bibfnamefont {A.~M.}\ \bibnamefont
  {Black-Schaffer}},\ }\bibfield  {title} {\bibinfo {title} {Highly tunable
  time-reversal-invariant topological superconductivity in topological
  insulator thin films},\ }\href@noop {} {\bibfield  {journal} {\bibinfo
  {journal} {Scientific reports}\ }\textbf {\bibinfo {volume} {7}},\ \bibinfo
  {pages} {1} (\bibinfo {year} {2017})}\BibitemShut {NoStop}%
\bibitem [{\citenamefont {Casas}\ \emph {et~al.}(2019)\citenamefont {Casas},
  \citenamefont {Arrachea}, \citenamefont {Herrera},\ and\ \citenamefont
  {Yeyati}}]{casas2019proximity}%
  \BibitemOpen
  \bibfield  {author} {\bibinfo {author} {\bibfnamefont {O.~E.}\ \bibnamefont
  {Casas}}, \bibinfo {author} {\bibfnamefont {L.}~\bibnamefont {Arrachea}},
  \bibinfo {author} {\bibfnamefont {W.~J.}\ \bibnamefont {Herrera}},\ and\
  \bibinfo {author} {\bibfnamefont {A.~L.}\ \bibnamefont {Yeyati}},\ }\bibfield
   {title} {\bibinfo {title} {Proximity induced time-reversal topological
  superconductivity in bi 2 se 3 films without phase tuning},\ }\href@noop {}
  {\bibfield  {journal} {\bibinfo  {journal} {Physical Review B}\ }\textbf
  {\bibinfo {volume} {99}},\ \bibinfo {pages} {161301(R)} (\bibinfo {year}
  {2019})}\BibitemShut {NoStop}%
\bibitem [{\citenamefont {Zhang}\ and\ \citenamefont
  {DasSarma}(2021)}]{zhang2021intrinsic}%
  \BibitemOpen
  \bibfield  {author} {\bibinfo {author} {\bibfnamefont {R.-X.}\ \bibnamefont
  {Zhang}}\ and\ \bibinfo {author} {\bibfnamefont {S.}~\bibnamefont
  {DasSarma}},\ }\bibfield  {title} {\bibinfo {title} {Intrinsic
  time-reversal-invariant topological superconductivity in thin films of
  iron-based superconductors},\ }\href@noop {} {\bibfield  {journal} {\bibinfo
  {journal} {Phys. Rev. Lett.}\ }\textbf {\bibinfo {volume} {126}},\ \bibinfo
  {pages} {137001} (\bibinfo {year} {2021})}\BibitemShut {NoStop}%
\bibitem [{\citenamefont {Chung}\ \emph {et~al.}(2013)\citenamefont {Chung},
  \citenamefont {Horowitz},\ and\ \citenamefont {Qi}}]{chung2013time}%
  \BibitemOpen
  \bibfield  {author} {\bibinfo {author} {\bibfnamefont {S.~B.}\ \bibnamefont
  {Chung}}, \bibinfo {author} {\bibfnamefont {J.}~\bibnamefont {Horowitz}},\
  and\ \bibinfo {author} {\bibfnamefont {X.-L.}\ \bibnamefont {Qi}},\
  }\bibfield  {title} {\bibinfo {title} {Time-reversal anomaly and josephson
  effect in time-reversal-invariant topological superconductors},\ }\href@noop
  {} {\bibfield  {journal} {\bibinfo  {journal} {Physical Review B}\ }\textbf
  {\bibinfo {volume} {88}},\ \bibinfo {pages} {214514} (\bibinfo {year}
  {2013})}\BibitemShut {NoStop}%
\bibitem [{\citenamefont {Nakosai}\ \emph {et~al.}(2013)\citenamefont
  {Nakosai}, \citenamefont {Budich}, \citenamefont {Tanaka}, \citenamefont
  {Trauzettel},\ and\ \citenamefont {Nagaosa}}]{nakosai2013majorana}%
  \BibitemOpen
  \bibfield  {author} {\bibinfo {author} {\bibfnamefont {S.}~\bibnamefont
  {Nakosai}}, \bibinfo {author} {\bibfnamefont {J.~C.}\ \bibnamefont {Budich}},
  \bibinfo {author} {\bibfnamefont {Y.}~\bibnamefont {Tanaka}}, \bibinfo
  {author} {\bibfnamefont {B.}~\bibnamefont {Trauzettel}},\ and\ \bibinfo
  {author} {\bibfnamefont {N.}~\bibnamefont {Nagaosa}},\ }\bibfield  {title}
  {\bibinfo {title} {Majorana bound states and nonlocal spin correlations in a
  quantum wire on an unconventional superconductor},\ }\href@noop {} {\bibfield
   {journal} {\bibinfo  {journal} {Phys. Rev. Lett.}\ }\textbf {\bibinfo
  {volume} {110}},\ \bibinfo {pages} {117002} (\bibinfo {year}
  {2013})}\BibitemShut {NoStop}%
\bibitem [{\citenamefont {Schrade}\ \emph {et~al.}(2015)\citenamefont
  {Schrade}, \citenamefont {Zyuzin}, \citenamefont {Klinovaja},\ and\
  \citenamefont {Loss}}]{schrade2015proximity}%
  \BibitemOpen
  \bibfield  {author} {\bibinfo {author} {\bibfnamefont {C.}~\bibnamefont
  {Schrade}}, \bibinfo {author} {\bibfnamefont {A.~A.}\ \bibnamefont {Zyuzin}},
  \bibinfo {author} {\bibfnamefont {J.}~\bibnamefont {Klinovaja}},\ and\
  \bibinfo {author} {\bibfnamefont {D.}~\bibnamefont {Loss}},\ }\bibfield
  {title} {\bibinfo {title} {Proximity-induced $\pi$ josephson junctions in
  topological insulators and kramers pairs of majorana fermions},\ }\href@noop
  {} {\bibfield  {journal} {\bibinfo  {journal} {Phys. Rev. Lett.}\ }\textbf
  {\bibinfo {volume} {115}},\ \bibinfo {pages} {237001} (\bibinfo {year}
  {2015})}\BibitemShut {NoStop}%
\bibitem [{\citenamefont {Li}\ \emph {et~al.}(2016)\citenamefont {Li},
  \citenamefont {Pan}, \citenamefont {Bernevig},\ and\ \citenamefont
  {Lutchyn}}]{li2016detection}%
  \BibitemOpen
  \bibfield  {author} {\bibinfo {author} {\bibfnamefont {J.}~\bibnamefont
  {Li}}, \bibinfo {author} {\bibfnamefont {W.}~\bibnamefont {Pan}}, \bibinfo
  {author} {\bibfnamefont {B.~A.}\ \bibnamefont {Bernevig}},\ and\ \bibinfo
  {author} {\bibfnamefont {R.~M.}\ \bibnamefont {Lutchyn}},\ }\bibfield
  {title} {\bibinfo {title} {Detection of majorana kramers pairs using a
  quantum point contact},\ }\href@noop {} {\bibfield  {journal} {\bibinfo
  {journal} {Phys. Rev. Lett.}\ }\textbf {\bibinfo {volume} {117}},\ \bibinfo
  {pages} {046804} (\bibinfo {year} {2016})}\BibitemShut {NoStop}%
\bibitem [{\citenamefont {Knapp}\ \emph {et~al.}(2020)\citenamefont {Knapp},
  \citenamefont {Chew},\ and\ \citenamefont {Alicea}}]{alicea}%
  \BibitemOpen
  \bibfield  {author} {\bibinfo {author} {\bibfnamefont {C.}~\bibnamefont
  {Knapp}}, \bibinfo {author} {\bibfnamefont {A.}~\bibnamefont {Chew}},\ and\
  \bibinfo {author} {\bibfnamefont {J.}~\bibnamefont {Alicea}},\ }\bibfield
  {title} {\bibinfo {title} {Fragility of the fractional josephson effect in
  time-reversal-invariant topological superconductors},\ }\href@noop {}
  {\bibfield  {journal} {\bibinfo  {journal} {Phys. Rev. Lett.}\ }\textbf
  {\bibinfo {volume} {125}},\ \bibinfo {pages} {207002} (\bibinfo {year}
  {2020})}\BibitemShut {NoStop}%
\bibitem [{\citenamefont {Camjayi}\ \emph {et~al.}(2017)\citenamefont
  {Camjayi}, \citenamefont {Arrachea}, \citenamefont {Aligia},\ and\
  \citenamefont {vonOppen}}]{camjayi2017fractional}%
  \BibitemOpen
  \bibfield  {author} {\bibinfo {author} {\bibfnamefont {A.}~\bibnamefont
  {Camjayi}}, \bibinfo {author} {\bibfnamefont {L.}~\bibnamefont {Arrachea}},
  \bibinfo {author} {\bibfnamefont {A.}~\bibnamefont {Aligia}},\ and\ \bibinfo
  {author} {\bibfnamefont {F.}~\bibnamefont {vonOppen}},\ }\bibfield  {title}
  {\bibinfo {title} {Fractional spin and josephson effect in
  time-reversal-invariant topological superconductors},\ }\href@noop {}
  {\bibfield  {journal} {\bibinfo  {journal} {Phys. Rev. Lett.}\ }\textbf
  {\bibinfo {volume} {119}},\ \bibinfo {pages} {046801} (\bibinfo {year}
  {2017})}\BibitemShut {NoStop}%
\bibitem [{\citenamefont {Schrade}\ and\ \citenamefont
  {Fu}(2018)}]{schrade2018parity}%
  \BibitemOpen
  \bibfield  {author} {\bibinfo {author} {\bibfnamefont {C.}~\bibnamefont
  {Schrade}}\ and\ \bibinfo {author} {\bibfnamefont {L.}~\bibnamefont {Fu}},\
  }\bibfield  {title} {\bibinfo {title} {Parity-controlled 2 $\pi$ josephson
  effect mediated by majorana kramers pairs},\ }\href@noop {} {\bibfield
  {journal} {\bibinfo  {journal} {Phys. Rev. Lett.}\ }\textbf {\bibinfo
  {volume} {120}},\ \bibinfo {pages} {267002} (\bibinfo {year}
  {2018})}\BibitemShut {NoStop}%
\bibitem [{\citenamefont {Aligia}\ and\ \citenamefont
  {Arrachea}(2018)}]{aligia2018entangled}%
  \BibitemOpen
  \bibfield  {author} {\bibinfo {author} {\bibfnamefont {A.~A.}\ \bibnamefont
  {Aligia}}\ and\ \bibinfo {author} {\bibfnamefont {L.}~\bibnamefont
  {Arrachea}},\ }\bibfield  {title} {\bibinfo {title} {Entangled end states
  with fractionalized spin projection in a time-reversal-invariant topological
  superconducting wire},\ }\href@noop {} {\bibfield  {journal} {\bibinfo
  {journal} {Physical Review B}\ }\textbf {\bibinfo {volume} {98}},\ \bibinfo
  {pages} {174507} (\bibinfo {year} {2018})}\BibitemShut {NoStop}%
\bibitem [{\citenamefont {Haim}\ and\ \citenamefont
  {Oreg}(2019)}]{haim2019time}%
  \BibitemOpen
  \bibfield  {author} {\bibinfo {author} {\bibfnamefont {A.}~\bibnamefont
  {Haim}}\ and\ \bibinfo {author} {\bibfnamefont {Y.}~\bibnamefont {Oreg}},\
  }\bibfield  {title} {\bibinfo {title} {Time-reversal-invariant topological
  superconductivity in one and two dimensions},\ }\href@noop {} {\bibfield
  {journal} {\bibinfo  {journal} {Physics Reports}\ }\textbf {\bibinfo {volume}
  {825}},\ \bibinfo {pages} {1} (\bibinfo {year} {2019})}\BibitemShut {NoStop}%
\bibitem [{\citenamefont {Gong}\ \emph {et~al.}(2016)\citenamefont {Gong},
  \citenamefont {Gao}, \citenamefont {Shan},\ and\ \citenamefont
  {Yi}}]{gong2016influence}%
  \BibitemOpen
  \bibfield  {author} {\bibinfo {author} {\bibfnamefont {W.-J.}\ \bibnamefont
  {Gong}}, \bibinfo {author} {\bibfnamefont {Z.}~\bibnamefont {Gao}}, \bibinfo
  {author} {\bibfnamefont {W.-F.}\ \bibnamefont {Shan}},\ and\ \bibinfo
  {author} {\bibfnamefont {G.-Y.}\ \bibnamefont {Yi}},\ }\bibfield  {title}
  {\bibinfo {title} {Influence of an embedded quantum dot on the josephson
  effect in the topological superconducting junction with majorana doublets},\
  }\href@noop {} {\bibfield  {journal} {\bibinfo  {journal} {Scientific
  reports}\ }\textbf {\bibinfo {volume} {6}},\ \bibinfo {pages} {1} (\bibinfo
  {year} {2016})}\BibitemShut {NoStop}%
\bibitem [{\citenamefont {Mashkoori}\ \emph {et~al.}(2019)\citenamefont
  {Mashkoori}, \citenamefont {Moghaddam}, \citenamefont {Hajibabaee},
  \citenamefont {Black-Schaffer},\ and\ \citenamefont
  {Parhizgar}}]{mashkoori2019impact}%
  \BibitemOpen
  \bibfield  {author} {\bibinfo {author} {\bibfnamefont {M.}~\bibnamefont
  {Mashkoori}}, \bibinfo {author} {\bibfnamefont {A.~G.}\ \bibnamefont
  {Moghaddam}}, \bibinfo {author} {\bibfnamefont {M.~H.}\ \bibnamefont
  {Hajibabaee}}, \bibinfo {author} {\bibfnamefont {A.~M.}\ \bibnamefont
  {Black-Schaffer}},\ and\ \bibinfo {author} {\bibfnamefont {F.}~\bibnamefont
  {Parhizgar}},\ }\bibfield  {title} {\bibinfo {title} {Impact of topology on
  the impurity effects in extended s-wave superconductors with spin-orbit
  coupling},\ }\href@noop {} {\bibfield  {journal} {\bibinfo  {journal}
  {Physical Review B}\ }\textbf {\bibinfo {volume} {99}},\ \bibinfo {pages}
  {014508} (\bibinfo {year} {2019})}\BibitemShut {NoStop}%
\bibitem [{\citenamefont {Lauke}\ \emph {et~al.}(2018)\citenamefont {Lauke},
  \citenamefont {Scheurer}, \citenamefont {Poenicke},\ and\ \citenamefont
  {Schmalian}}]{lauke2018friedel}%
  \BibitemOpen
  \bibfield  {author} {\bibinfo {author} {\bibfnamefont {L.}~\bibnamefont
  {Lauke}}, \bibinfo {author} {\bibfnamefont {M.~S.}\ \bibnamefont {Scheurer}},
  \bibinfo {author} {\bibfnamefont {A.}~\bibnamefont {Poenicke}},\ and\
  \bibinfo {author} {\bibfnamefont {J.}~\bibnamefont {Schmalian}},\ }\bibfield
  {title} {\bibinfo {title} {Friedel oscillations and majorana zero modes in
  inhomogeneous superconductors},\ }\href@noop {} {\bibfield  {journal}
  {\bibinfo  {journal} {Physical Review B}\ }\textbf {\bibinfo {volume} {98}},\
  \bibinfo {pages} {134502} (\bibinfo {year} {2018})}\BibitemShut {NoStop}%
\bibitem [{\citenamefont {Arrachea}\ \emph {et~al.}(2019)\citenamefont
  {Arrachea}, \citenamefont {Camjayi}, \citenamefont {Aligia},\ and\
  \citenamefont {Gru{\~n}eiro}}]{arrachea2019catalog}%
  \BibitemOpen
  \bibfield  {author} {\bibinfo {author} {\bibfnamefont {L.}~\bibnamefont
  {Arrachea}}, \bibinfo {author} {\bibfnamefont {A.}~\bibnamefont {Camjayi}},
  \bibinfo {author} {\bibfnamefont {A.~A.}\ \bibnamefont {Aligia}},\ and\
  \bibinfo {author} {\bibfnamefont {L.}~\bibnamefont {Gru{\~n}eiro}},\
  }\bibfield  {title} {\bibinfo {title} {Catalog of andreev spectra and
  josephson effects in structures with time-reversal-invariant topological
  superconductor wires},\ }\href@noop {} {\bibfield  {journal} {\bibinfo
  {journal} {Physical Review B}\ }\textbf {\bibinfo {volume} {99}},\ \bibinfo
  {pages} {085431} (\bibinfo {year} {2019})}\BibitemShut {NoStop}%
\bibitem [{\citenamefont {Haim}(2019)}]{haim2019spontaneous}%
  \BibitemOpen
  \bibfield  {author} {\bibinfo {author} {\bibfnamefont {A.}~\bibnamefont
  {Haim}},\ }\bibfield  {title} {\bibinfo {title} {Spontaneous josephson $\pi$
  junctions with topological superconductors},\ }\href@noop {} {\bibfield
  {journal} {\bibinfo  {journal} {Physical Review B}\ }\textbf {\bibinfo
  {volume} {100}},\ \bibinfo {pages} {064505} (\bibinfo {year}
  {2019})}\BibitemShut {NoStop}%
\bibitem [{\citenamefont {Matano}\ \emph {et~al.}(2016)\citenamefont {Matano},
  \citenamefont {Kriener}, \citenamefont {Segawa}, \citenamefont {Ando},\ and\
  \citenamefont {Zheng}}]{matano2016spin}%
  \BibitemOpen
  \bibfield  {author} {\bibinfo {author} {\bibfnamefont {K.}~\bibnamefont
  {Matano}}, \bibinfo {author} {\bibfnamefont {M.}~\bibnamefont {Kriener}},
  \bibinfo {author} {\bibfnamefont {K.}~\bibnamefont {Segawa}}, \bibinfo
  {author} {\bibfnamefont {Y.}~\bibnamefont {Ando}},\ and\ \bibinfo {author}
  {\bibfnamefont {G.-q.}\ \bibnamefont {Zheng}},\ }\bibfield  {title} {\bibinfo
  {title} {Spin-rotation symmetry breaking in the superconducting state of
  cuxbi2se3},\ }\href@noop {} {\bibfield  {journal} {\bibinfo  {journal}
  {Nature Physics}\ }\textbf {\bibinfo {volume} {12}},\ \bibinfo {pages} {852}
  (\bibinfo {year} {2016})}\BibitemShut {NoStop}%
\bibitem [{\citenamefont {Yonezawa}\ \emph {et~al.}(2017)\citenamefont
  {Yonezawa}, \citenamefont {Tajiri}, \citenamefont {Nakata}, \citenamefont
  {Nagai}, \citenamefont {Wang}, \citenamefont {Segawa}, \citenamefont {Ando},\
  and\ \citenamefont {Maeno}}]{yonezawa2017thermodynamic}%
  \BibitemOpen
  \bibfield  {author} {\bibinfo {author} {\bibfnamefont {S.}~\bibnamefont
  {Yonezawa}}, \bibinfo {author} {\bibfnamefont {K.}~\bibnamefont {Tajiri}},
  \bibinfo {author} {\bibfnamefont {S.}~\bibnamefont {Nakata}}, \bibinfo
  {author} {\bibfnamefont {Y.}~\bibnamefont {Nagai}}, \bibinfo {author}
  {\bibfnamefont {Z.}~\bibnamefont {Wang}}, \bibinfo {author} {\bibfnamefont
  {K.}~\bibnamefont {Segawa}}, \bibinfo {author} {\bibfnamefont
  {Y.}~\bibnamefont {Ando}},\ and\ \bibinfo {author} {\bibfnamefont
  {Y.}~\bibnamefont {Maeno}},\ }\bibfield  {title} {\bibinfo {title}
  {Thermodynamic evidence for nematic superconductivity in cuxbi2se3},\
  }\href@noop {} {\bibfield  {journal} {\bibinfo  {journal} {Nature Physics}\
  }\textbf {\bibinfo {volume} {13}},\ \bibinfo {pages} {123} (\bibinfo {year}
  {2017})}\BibitemShut {NoStop}%
\bibitem [{\citenamefont {Vollhardt}\ and\ \citenamefont
  {Wolfle}(2013)}]{vollhardt2013superfluid}%
  \BibitemOpen
  \bibfield  {author} {\bibinfo {author} {\bibfnamefont {D.}~\bibnamefont
  {Vollhardt}}\ and\ \bibinfo {author} {\bibfnamefont {P.}~\bibnamefont
  {Wolfle}},\ }\href@noop {} {\emph {\bibinfo {title} {The superfluid phases of
  helium 3}}}\ (\bibinfo  {publisher} {Courier Corporation},\ \bibinfo {year}
  {2013})\BibitemShut {NoStop}%
\bibitem [{\citenamefont {Read}\ and\ \citenamefont
  {Green}(2000)}]{read2000paired}%
  \BibitemOpen
  \bibfield  {author} {\bibinfo {author} {\bibfnamefont {N.}~\bibnamefont
  {Read}}\ and\ \bibinfo {author} {\bibfnamefont {D.}~\bibnamefont {Green}},\
  }\bibfield  {title} {\bibinfo {title} {Paired states of fermions in two
  dimensions with breaking of parity and time-reversal symmetries and the
  fractional quantum hall effect},\ }\href@noop {} {\bibfield  {journal}
  {\bibinfo  {journal} {Physical Review B}\ }\textbf {\bibinfo {volume} {61}},\
  \bibinfo {pages} {10267} (\bibinfo {year} {2000})}\BibitemShut {NoStop}%
\bibitem [{\citenamefont {Aligia}\ \emph {et~al.}(2020)\citenamefont {Aligia},
  \citenamefont {PerezDaroca},\ and\ \citenamefont
  {Arrachea}}]{aligia2020tomography}%
  \BibitemOpen
  \bibfield  {author} {\bibinfo {author} {\bibfnamefont {A.~A.}\ \bibnamefont
  {Aligia}}, \bibinfo {author} {\bibfnamefont {D.}~\bibnamefont
  {PerezDaroca}},\ and\ \bibinfo {author} {\bibfnamefont {L.}~\bibnamefont
  {Arrachea}},\ }\bibfield  {title} {\bibinfo {title} {Tomography of
  zero-energy end modes in topological superconducting wires},\ }\href@noop {}
  {\bibfield  {journal} {\bibinfo  {journal} {Phys. Rev. Lett.}\ }\textbf
  {\bibinfo {volume} {125}},\ \bibinfo {pages} {256801} (\bibinfo {year}
  {2020})}\BibitemShut {NoStop}%
\bibitem [{\citenamefont {Alase}\ \emph {et~al.}(2016)\citenamefont {Alase},
  \citenamefont {Cobanera}, \citenamefont {Ortiz},\ and\ \citenamefont
  {Viola}}]{alaseprl}%
  \BibitemOpen
  \bibfield  {author} {\bibinfo {author} {\bibfnamefont {A.}~\bibnamefont
  {Alase}}, \bibinfo {author} {\bibfnamefont {E.}~\bibnamefont {Cobanera}},
  \bibinfo {author} {\bibfnamefont {G.}~\bibnamefont {Ortiz}},\ and\ \bibinfo
  {author} {\bibfnamefont {L.}~\bibnamefont {Viola}},\ }\bibfield  {title}
  {\bibinfo {title} {Exact solution of quadratic fermionic hamiltonians for
  arbitrary boundary conditions},\ }\href
  {https://doi.org/10.1103/PhysRevLett.117.076804} {\bibfield  {journal}
  {\bibinfo  {journal} {Phys. Rev. Lett.}\ }\textbf {\bibinfo {volume} {117}},\
  \bibinfo {pages} {076804} (\bibinfo {year} {2016})}\BibitemShut {NoStop}%
\bibitem [{\citenamefont {Alase}\ \emph {et~al.}(2017)\citenamefont {Alase},
  \citenamefont {Cobanera}, \citenamefont {Ortiz},\ and\ \citenamefont
  {Viola}}]{alaseprb}%
  \BibitemOpen
  \bibfield  {author} {\bibinfo {author} {\bibfnamefont {A.}~\bibnamefont
  {Alase}}, \bibinfo {author} {\bibfnamefont {E.}~\bibnamefont {Cobanera}},
  \bibinfo {author} {\bibfnamefont {G.}~\bibnamefont {Ortiz}},\ and\ \bibinfo
  {author} {\bibfnamefont {L.}~\bibnamefont {Viola}},\ }\bibfield  {title}
  {\bibinfo {title} {Generalization of bloch's theorem for arbitrary boundary
  conditions: Theory},\ }\href {https://doi.org/10.1103/PhysRevB.96.195133}
  {\bibfield  {journal} {\bibinfo  {journal} {Phys. Rev. B}\ }\textbf {\bibinfo
  {volume} {96}},\ \bibinfo {pages} {195133} (\bibinfo {year}
  {2017})}\BibitemShut {NoStop}%
\bibitem [{\citenamefont {Aligia}\ and\ \citenamefont
  {Camjayi}(2019)}]{aligia2019}%
  \BibitemOpen
  \bibfield  {author} {\bibinfo {author} {\bibfnamefont {A.~A.}\ \bibnamefont
  {Aligia}}\ and\ \bibinfo {author} {\bibfnamefont {A.}~\bibnamefont
  {Camjayi}},\ }\bibfield  {title} {\bibinfo {title} {Exact analytical solution
  of a time-reversal-invariant topological superconducting wire},\ }\href@noop
  {} {\bibfield  {journal} {\bibinfo  {journal} {Phys. Rev. B}\ }\textbf
  {\bibinfo {volume} {100}},\ \bibinfo {pages} {115413} (\bibinfo {year}
  {2019})}\BibitemShut {NoStop}%
\bibitem [{\citenamefont {Tanaka}\ and\ \citenamefont
  {Kashiwaya}(1996)}]{tanaka1}%
  \BibitemOpen
  \bibfield  {author} {\bibinfo {author} {\bibfnamefont {Y.}~\bibnamefont
  {Tanaka}}\ and\ \bibinfo {author} {\bibfnamefont {S.}~\bibnamefont
  {Kashiwaya}},\ }\bibfield  {title} {\bibinfo {title} {Theory of the josephson
  effect in d-wave superconductors},\ }\href@noop {} {\bibfield  {journal}
  {\bibinfo  {journal} {Physical Review B}\ }\textbf {\bibinfo {volume} {53}},\
  \bibinfo {pages} {R11957} (\bibinfo {year} {1996})}\BibitemShut {NoStop}%
\bibitem [{\citenamefont {Tanaka}\ and\ \citenamefont
  {Kashiwaya}(1997)}]{tanaka2}%
  \BibitemOpen
  \bibfield  {author} {\bibinfo {author} {\bibfnamefont {Y.}~\bibnamefont
  {Tanaka}}\ and\ \bibinfo {author} {\bibfnamefont {S.}~\bibnamefont
  {Kashiwaya}},\ }\bibfield  {title} {\bibinfo {title} {Theory of josephson
  effects in anisotropic superconductors},\ }\href
  {https://doi.org/10.1103/PhysRevB.56.892} {\bibfield  {journal} {\bibinfo
  {journal} {Phys. Rev. B}\ }\textbf {\bibinfo {volume} {56}},\ \bibinfo
  {pages} {892} (\bibinfo {year} {1997})}\BibitemShut {NoStop}%
\bibitem [{\citenamefont {Kashiwaya}\ and\ \citenamefont
  {Tanaka}(2000)}]{tanaka3}%
  \BibitemOpen
  \bibfield  {author} {\bibinfo {author} {\bibfnamefont {S.}~\bibnamefont
  {Kashiwaya}}\ and\ \bibinfo {author} {\bibfnamefont {Y.}~\bibnamefont
  {Tanaka}},\ }\bibfield  {title} {\bibinfo {title} {Tunnelling effects on
  surface bound states in unconventional superconductors},\ }\href
  {https://doi.org/10.1088/0034-4885/63/10/202} {\bibfield  {journal} {\bibinfo
   {journal} {Reports on Progress in Physics}\ }\textbf {\bibinfo {volume}
  {63}},\ \bibinfo {pages} {1641} (\bibinfo {year} {2000})}\BibitemShut
  {NoStop}%
\bibitem [{\citenamefont {Alidoust}\ \emph {et~al.}(2018)\citenamefont
  {Alidoust}, \citenamefont {Willatzen},\ and\ \citenamefont
  {Jauho}}]{alidoust2018strain}%
  \BibitemOpen
  \bibfield  {author} {\bibinfo {author} {\bibfnamefont {M.}~\bibnamefont
  {Alidoust}}, \bibinfo {author} {\bibfnamefont {M.}~\bibnamefont
  {Willatzen}},\ and\ \bibinfo {author} {\bibfnamefont {A.-P.}\ \bibnamefont
  {Jauho}},\ }\bibfield  {title} {\bibinfo {title} {Strain-engineered majorana
  zero energy modes and $\varphi$ 0 josephson state in black phosphorus},\
  }\href@noop {} {\bibfield  {journal} {\bibinfo  {journal} {Physical Review
  B}\ }\textbf {\bibinfo {volume} {98}},\ \bibinfo {pages} {085414} (\bibinfo
  {year} {2018})}\BibitemShut {NoStop}%
\bibitem [{\citenamefont {Alidoust}(2020)}]{alidoust2020critical}%
  \BibitemOpen
  \bibfield  {author} {\bibinfo {author} {\bibfnamefont {M.}~\bibnamefont
  {Alidoust}},\ }\bibfield  {title} {\bibinfo {title} {Critical supercurrent
  and $\varphi$ 0 state for probing a persistent spin helix},\ }\href@noop {}
  {\bibfield  {journal} {\bibinfo  {journal} {Physical Review B}\ }\textbf
  {\bibinfo {volume} {101}},\ \bibinfo {pages} {155123} (\bibinfo {year}
  {2020})}\BibitemShut {NoStop}%
\bibitem [{\citenamefont {Alidoust}\ \emph {et~al.}(2021)\citenamefont
  {Alidoust}, \citenamefont {Shen},\ and\ \citenamefont
  {{\v{Z}}uti{\'c}}}]{alidoust2021cubic}%
  \BibitemOpen
  \bibfield  {author} {\bibinfo {author} {\bibfnamefont {M.}~\bibnamefont
  {Alidoust}}, \bibinfo {author} {\bibfnamefont {C.}~\bibnamefont {Shen}},\
  and\ \bibinfo {author} {\bibfnamefont {I.}~\bibnamefont {{\v{Z}}uti{\'c}}},\
  }\bibfield  {title} {\bibinfo {title} {Cubic spin-orbit coupling and
  anomalous josephson effect in planar junctions},\ }\href@noop {} {\bibfield
  {journal} {\bibinfo  {journal} {Physical Review B}\ }\textbf {\bibinfo
  {volume} {103}},\ \bibinfo {pages} {L060503} (\bibinfo {year}
  {2021})}\BibitemShut {NoStop}%
\bibitem [{\citenamefont {Zazunov}\ \emph {et~al.}(2009)\citenamefont
  {Zazunov}, \citenamefont {Egger}, \citenamefont {Jonckheere},\ and\
  \citenamefont {Martin}}]{zazunov2009anomalous}%
  \BibitemOpen
  \bibfield  {author} {\bibinfo {author} {\bibfnamefont {A.}~\bibnamefont
  {Zazunov}}, \bibinfo {author} {\bibfnamefont {R.}~\bibnamefont {Egger}},
  \bibinfo {author} {\bibfnamefont {T.}~\bibnamefont {Jonckheere}},\ and\
  \bibinfo {author} {\bibfnamefont {T.}~\bibnamefont {Martin}},\ }\bibfield
  {title} {\bibinfo {title} {Anomalous josephson current through a spin-orbit
  coupled quantum dot},\ }\href@noop {} {\bibfield  {journal} {\bibinfo
  {journal} {Physical review letters}\ }\textbf {\bibinfo {volume} {103}},\
  \bibinfo {pages} {147004} (\bibinfo {year} {2009})}\BibitemShut {NoStop}%
\end{thebibliography}%

%

\end{document}